# Exoplanet Atmospheres

## Sara Seager (MIT) and Drake Deming (NASA/GSFC)

## Abstract

At the dawn of the first discovery of exoplanets orbiting sun-like stars in the mid-1990s, few believed that observations of exoplanet atmospheres would ever be possible. After the 2002 *Hubble Space Telescope* detection of a transiting exoplanet atmosphere, many skeptics discounted it as a one-object, one-method success. Nevertheless, the field is now firmly established, with over two dozen exoplanet atmospheres observed today. Hot Jupiters are the type of exoplanet currently most amenable to study. Highlights include: detection of molecular spectral features; observation of day-night temperature gradients; and constraints on vertical atmospheric structure. Atmospheres of giant planets far from their host stars are also being studied with direct imaging. The ultimate exoplanet goal is to answer the enigmatic and ancient question, "Are we alone?" via detection of atmospheric biosignatures. Two exciting prospects are the immediate focus on transiting super Earths orbiting in the habitable zone of M-dwarfs, and ultimately the spaceborne direct imaging of true Earth analogs.

## 1. Introduction

Exoplanets are unique objects in astronomy because they have local counterparts—the solar system planets—available for study. To introduce the field of exoplanet atmospheres we therefore begin with a brief description of the relevant history of Solar System planet atmospheres before turning to a concise history of exoplanet atmosphere research.

### 1.1 Solar and Extrasolar Planets

Atmospheres surrounding planets in our solar system have been known and studied since the 19th century (Challis, 1863). Early solar system observers noted that satellites and stars disappear gradually, not instantaneously, when occulted by the planet. They observed variable features on the planets that did not change on a regular cycle, as would surface features on a rotating object. They recognized that these properties proved the existence of atmospheres. However, the first spectroscopic observations of atmospheres on solar system planets revealed—to the surprise of many—that these atmospheres were very unlike Earth's. As late as the 1920s astronomers were amazed to find that the atmosphere of Venus did *not* contain oxygen (Webster, 1927). Our understanding of planetary atmospheres developed in close parallel with the continued application of spectroscopy. Early spectroscopic successes include the identification of methane in the atmospheres of the giant planets (Adel & Slipher, 1934), carbon dioxide on the terrestrial planets (Adel,1937), and the spectroscopic detection of an atmosphere on Titan (Kuiper, 1944). As a result of these early observations, the basic physics and chemistry of planetary atmospheres was established.

While atmospheres in the solar system are scientifically interesting in their own right, by the middle- to late-20$^{th}$ century, another major motivation developed for studying solar system atmospheres in intricate detail. This was to facilitate the orbiting and landing of spacecraft on solar system planets, for example the spectacular aerobraking and orbit insertion of the Mars Reconnaisance Orbiter (Zurek & Smrekar, 2007).

Exoplanets have a similar driver larger than pure scientific curiosity. The ultimate goal for many researchers is the search for habitable exoplanets. The exoplanet atmosphere is the *only* way to infer whether or not a planet is habitable or likely inhabited; the planetary atmosphere is our window into temperatures, habitability indicators, and biosignature gases.

Exoplanet science has benefitted tremendously from the decades of work on solar system planets. No other field in astronomy has a pool of local analogs with highly detailed observations and a long established theoretical foundation. Nevertheless, one incontrovertible difference between solar and extrasolar planets will always remain: solar system planets are brighter and observable to much higher signal-to-noise levels than exoplanets. From the start, solar system planets were always manifestly bright, and their observations were never photon-starved. The field of exoplanet atmospheric studies, therefore, is not just one of extending old physics and chemistry to new types of planets, but is a research area of extremely challenging observations and development of new observational techniques.

## 1.2 A Brief History of Exoplanet Atmospheres

The dawn of the discovery of exoplanets orbiting sun-like stars took place in the mid 1990s, when radial velocity detections began and accelerated. Because of detection selection effects, many of the exoplanets found in the first few years of discovery orbited exceedingly close to their host star. Dubbed hot Jupiters, these planets orbit many times closer to their star than Mercury does to our sun. With semi-major axes ≤ 0.05 AU, the hot Jupiters are heated externally by their stars to temperatures of 1000—2000 K, or even higher. From the start the high temperature and close stellar proximity of hot Jupiters were recognized as favorable for atmospheric detection (Seager & Sasselov 1998).

Surprised by the implicit challenge to the solar system paradigm, some astronomers resisted the new exoplanet detections. The skeptics focused on a new, unknown type of stellar pulsation to explain the Doppler wobble, evidenced by possible spectral line asymmetries (e.g., Gray, 1997). Eventually, enough planets were discovered too far from their host stars to be explained away as stellar pulsations. Even with the debate about the exoplanet detections winding down in the late 1990s, few thought that exoplanet atmospheres could be observed at any time in the foreseeable future.

As the numbers of short-period exoplanets was rising (just under 30 by the end of the 20[th] century[1]) so too was the anticipation for the discovery of a transiting planet. A transiting planet is one that passes in front of the parent star as seen from Earth. A transit signature consistent with the Doppler phase would be incontrovertible evidence for an exoplanet. With a probability to transit of $R_*/a$, where $R_*$ is the stellar radius and $a$ is the semi-major axis, each hot Jupiter has about a 10% chance to transit. By the time about seven hot Jupiters were known, one of us started writing a paper on transit transmission spectra as a way to identify atomic and molecular features in exoplanet atmospheres, with a focus on atomic sodium (Seager & Sasselov 2000). HD 209458b was found to show transits at the end of 1999 (Charbonneau et al. 2000; Henry et al. 2000), and the first detection of an exoplanet atmosphere, via atomic sodium, with the *Hubble Space Telescope* soon followed (Charbonneau et al. 2002).

The excitement and breakthrough of the first exoplanet atmosphere detection was damped in the wider astronomy community in two ways. First, the sodium detection was at the 4.1 σ level (Charbonneau et al. 2002), and despite the thorough statistical tests carried out to support the detection, many in the community accustomed to much higher signal-to-noise (S/N) observations were wary. Second, those that did embrace the first exoplanet atmosphere discovery challenged it as a one-object, one-method success, because no other transiting planets were known or seemed on the horizon.

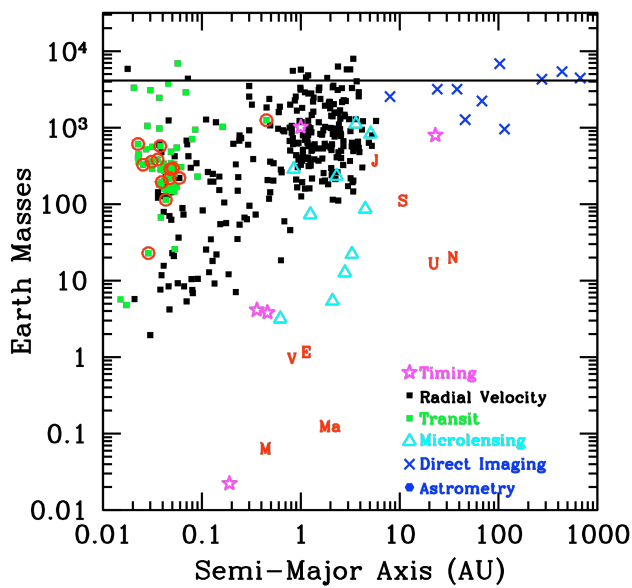

Figure 1. Known planets as of January 2010. Red letters indicate solar system planets. The red circles represent planets with published atmosphere detections. The solid line is the conventional upper mass limit for the definition of a planet. Data taken from http://exoplanet.eu/.

The theory of exoplanet atmospheres was also developing during the same time period, with several different thrusts. At that time, theory was leading observation, and observers consulted the model predictions to help define the most promising detection techniques. Most theory papers focused on irradiated hot Jupiters, emphasizing altered 1D temperature/pressure profiles resulting from the intense external irradiation by the host star, as well as the

---

effects of molecules such as water vapor (Seager & Sasselov 1998; Marley et al. 1999; Sudarsky, Burrows, & Pinto 2000; Barman, Hauschildt & Allard 2001), and borrowing from brown dwarf observations and models (Oppenheimer et al. 1998). Work on cloud modeling (Ackerman and Marley 2001; Cooper et al. 2003) and atmospheric circulation (Showman & Guillot 2002; Cho et al. 2003) followed soon thereafter. Calculation of exoplanet illumination phase curves, polarization curves (Seager, Whitney, & Sasselov 2000), and especially transmission spectra (Seager & Sasselov 2000; Brown 2001; Hubbard et al. 2001) set the stage for observed spectroscopy during transit. Taken together, this set of pioneering work built the foundation for the subsequent detection and study of exoplanet atmospheres at that early time when theory was leading observation.

In 2002, we and others began planning for secondary eclipse measurements of transiting hot Jupiters using the *Spitzer Space Telescope*, launched in August 2003. At mid-infrared wavelengths, hot Jupiters have a high planet-to-star contrast ratio, and the star and planet typically are bright enough to allow high precision photon-limited measurements. We expected to measure the planetary brightness temperature at thermal wavelengths, as the planet disappeared and reemerged from behind the host star. Coincidentally, other transiting planets were beginning to be found (Konacki et al. 2003). Although the first secondary eclipse detections (Charbonneau et al. 2005; Deming et al. 2005a) were statistically robust at the just under 6-$\sigma$ level, they again were not absolutely convincing to the entire astronomy community. Any remaining doubts, however, vanished with *Spitzer*'s 16 $\mu$m secondary eclipse observation of HD 189733 (Deming et al. 2006). That measurement showed an obvious eclipse, with amplitude 40 times the error level. This unleashed a flood of secondary eclipse observational detections using *Spitzer*. Today we can count secondary eclipses of about a dozen planets successfully observed by *Spitzer* at several wavelengths[2], with results published (Figure 1 and 2). An additional two dozen hot Jupiters have been observed by *Spitzer*, with results under analysis, or have observations planned by Warm *Spitzer* (see Table 1). It is accurate to say that no one anticipated the full magnitude and stunning impact of the *Spitzer Space Telescope* as a tool to develop the field of exoplanet atmospheric studies.

From its lonely beginnings just over a decade ago, the pendulum has swung to the point where exoplanet atmospheric researchers now populate a full-fledged field, and have produced over one hundred published papers. Today we can count atmospheric observations of dozens of exoplanets, and particularly note observations of 8 especially significant exoplanets (Figure 3). Skeptics are held at bay by the monumental and pioneering achievements of the first decade of exoplanet atmospheric research. Indeed the huge promise for the future is based on the incredible achievements of the past decade. This review (as of January 2010) takes a critical, albeit not exhaustive, look at the discoveries and future potential of exoplanet atmosphere studies.

---

[2] http://ssc.spitzer.caltech.edu/approvdprog/

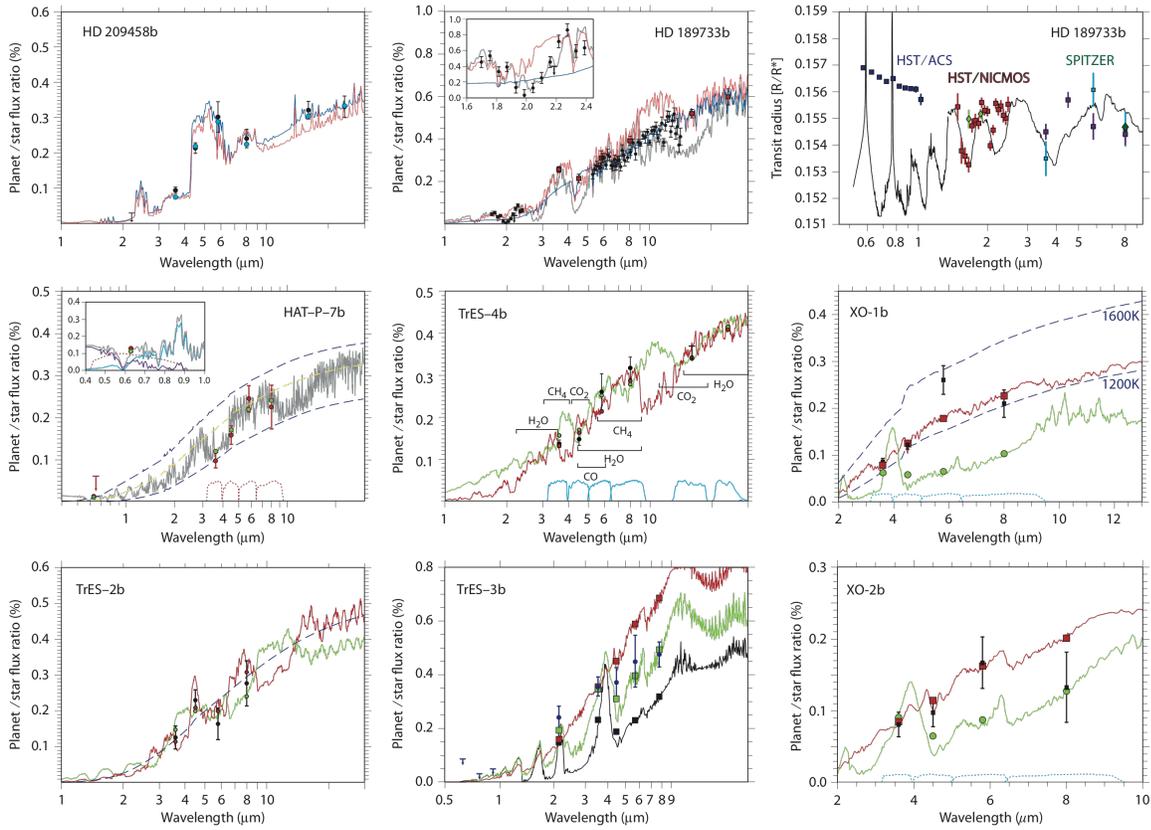

Figure 2. Collage of exoplanet atmosphere data and measurements. Most observed exoplanet atmospheres are the four *Spitzer*/IRAC bands via secondary eclipse. For references see Table 1.

Table 1. *Spitzer* IRAC Broad-Band Photometry

| 10 | 3.6 μm | 4.5 μm | 5.8 μm | 8 μm | Reference |
|---|---|---|---|---|---|
| HD189733b | 0.256% ±0.014% | 0.214% ±0.020% | 0.310% ±0.034% | 0.391 ±0.022% | Charbonneau et al. 2008 |
| HD209458b | 0.094% ±0.009% | 0.213% ±0.015% | 0.301% ±0.043% | 0.240% ±0.026% | Knutson et al. 2008 |
| HD149026b | XX | XX | XX | 0.084% +0.006% - 0.012% | Harrington et al. 2007 |
| HD80606b | | XX | | 0.0411% ± 0.0076% | Knutson et al. 2009c |
| | | | | 0.136% ±0.018% | Laughlin et al. 2009 |
| GJ436b | XX | XX | XX | 0.057% ± 0.008% | Deming et al. 2007 |
| | | | | 0.054% ± 0.007% | Demory et al. 2007 |
| CoRoT-1 | XX | XX | | | |
| CoRoT-2 | XX | 0.510% ±0.042% | | 0.41 ±0.11 % | Gillon et al. 2010 |
| HAT-1 | 0.080% ± 0.008% | 0.135% ± 0.022% | 0.203% ± 0.031% | 0.238% ±0.040% | Todorov et al. 2010 |
| HAT-2 | X | X | XX | XX | |
| HAT-3 | X | X | | | |
| HAT-4 | X | X | | | |
| HAT-5 | XX | XX | | | |
| HAT-6 | X | X | | | |
| HAT-7 | 0.098% ±0.017% | 0.159% ±0.022% | 0.245% ±0.031% | 0.225% ±0.052% | Christiansen et al. 2010 |
| HAT-8 | X | XX | | | |
| HAT-10 | X | XX | | | |
| HAT-11 | X | X | | | |
| HAT-12 | X | X | | | |
| TrES-1 | XX | 0.066% ± 0.013% | XX | 0.225% ±0.036% | Charbonneau et al. 2005 |
| TrES-2 | 0.135% ± 0.036% | 0.245% ± 0.027%, | 0.162% ± 0.064%, | 0.295% ±0.066%, | O'Donovan et al. 2010 |
| TrES-3 | 0.346% ±0.035% | 0.372% ±0.054% | 0.449% ±0.097% | 0.475% ±0.046% | Fressin et al. 2010 |
| TrES-4 | 0.137% ±0.011% | 0.148% ±0.016% | 0.261% ±0.059% | 0.318% ±0.044% | Knutson et al. 2009a |
| WASP-1 | XX | XX | XX | XX | |
| WASP-2 | XX | XX | XX | | |
| WASP-3 | XX | XX | | XX | |
| WASP-4 | XX | XX | | | |
| WASP-5 | XX | X | | | |
| WASP-6 | XX | X | | | |
| WASP-7 | X | X | | | |
| WASP-8 | | XX | | XX | |
| WASP-10 | X | X | | | |
| WASP-12 | XX | XX | XX | XX | |
| WASP-14 | X | XX | | XX | |

| | | | | |
|---|---|---|---|---|
| WASP-17 | | XX | | XX |
| WASP-18 | XX | XX | XX | XX |
| WASP-19 | XX | XX | XX | XX |
| XO-1 | 0.086% ±0.007% | 0.122% ±0.009% | 0.261% ±0.031% | 0.210% ±0.029% | Machalek et al. 2008 |
| | 0.081% ± | 0.098% ± | 0.167% ± | | |
| XO-2 | 0.017% | 0.020% | 0.036% | 0.133% ± 0.049% | Machalek et al. 2009 |
| XO-3 | 0.101% ±0.004% | 0.143% ±0.006% | 0.134% ±0.049% | 0.150% ±0.036% | Machalek et al. 2010 |
| XO-4 | XX | XX | X | | |

<u>Table 1.</u>  Tabulation of the exoplanet secondary eclipse observations.  Reported values are published measurements.  A double x refers to data in hand and analyses under way.  A single x refers to observations officially planned by *Spitzer* as of January 2010.  Wavelengths are in μm.

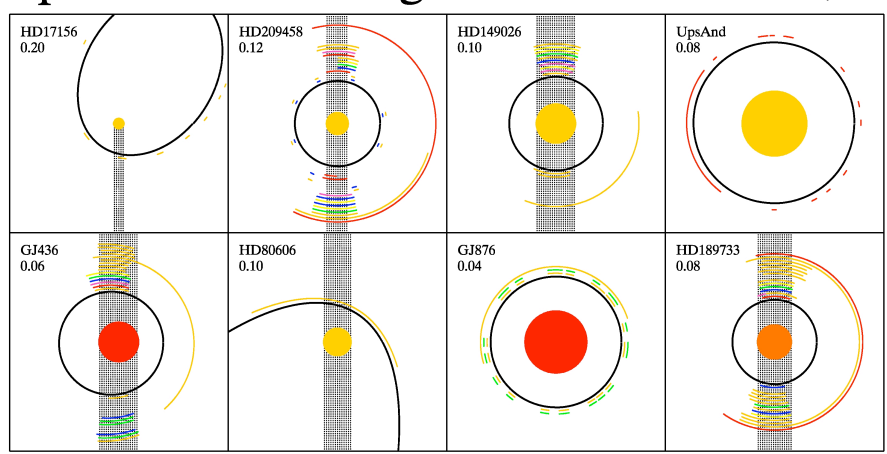

Figure 3.  Panel illustrating *Spitzer* photometry for eight exoplanets including some of the most interesting and significant exoplanets.  Each sub-panel has a different spatial scale, listed in AU below the planet name.  The *Spitzer* observational periods are indicated by colored arcs, with color indicating wavelength (see legend at top).  The very short arcs that appear similar to dots are short-duration *Spitzer* observations.  The spectral type of the star (late-F/G, K, or M) is indicated by color, and the stellar radii have been increased by a factor of 3 for all systems, for greater visibility.  Shading indicates the presence of a transit (lower shaded region),

## 2. Overview of Exoplanet Atmosphere Observations and Models

As a continued introduction we turn to background material for understanding exoplanet atmosphere observations and models.

### 2.1 Observations

### 2.1.1 Direct Imaging

The most natural way to think of observing exoplanet atmospheres is by taking an image of the exoplanet. This so-called "direct imaging" of planets is currently limited to big, bright, young or massive planets located far from their stars (see Figure 1). Direct imaging of substellar objects is currently possible with large ground-based telescopes and adaptive optics to cancel the atmosphere's blurring effects. Out of a dozen planet candidates, the most definitive planet detections are Fomalhaut b because of its mass (≤ 3 M$_J$) (Kalas et al. 2008) and the three planets orbiting HR 8799  (Marois et al. 2008). Not only do the HR 8799 planets have mass estimates below the brown dwarf limit, but the hierarchy of three objects orbiting a central star is simply not seen for multiple star systems.

Solar-system-like small exoplanets are not observable via direct imaging with current technology, even though an Earth at 10 pc is brighter than the faintest galaxies observed by the *Hubble Space Telescope* (*HST*). The major impediment to direct imaging of exoEarths is instead the adjacent host star; the sun is 10 million to 10 billion times brighter than Earth (for mid-infrared and visible wavelengths, respectively).  No existing or planned telescope is capable of achieving this contrast ratio at 1 AU separations. The current state of the art HR 8799 observations detected a planet at a contrast of 1/100,000 at a separation of about 0.5 arcsec. Fortunately much research and technology development is ongoing to enable space-based direct imaging of solar system aged Earths and Jupiters in the future.  See

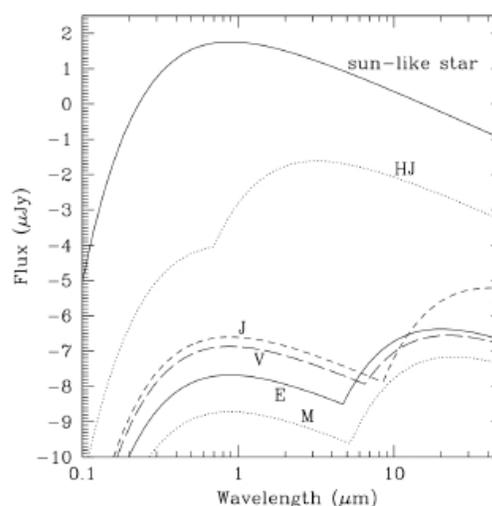

Figure 4. Black body flux (in units of 10$^{-26}$ W m$^{-2}$ Hz$^{-1}$) of some solar system bodies as "seen" from 10 pc. The Sun is represented by a 5750 K black body. The planets Jupiter, Venus, Earth, and Mars are shown and are labeled with their first initial.  A putative hot Jupiter is labeled with "HJ".  The planets have two peaks in their spectra. The short-wavelength peak is due to sunlight scattered from the planet atmosphere and is computed using the planet's geometric albedo. The long-wavelength peak is from the planet's thermal emission and is estimated by a black body of the planet's effective temperature. The hot Jupiter albedo was assumed to be 0.05 and the equilibrium temperature to be 1600 K. Temperature and albedo data was taken from Cox (2000).

Figure 4 for estimates of planetary fluxes, and Seager (2010; Chapter 3) for approximate formulae for order of magnitude estimates for direct imaging.

## 2.1.2 Transiting Exoplanet Atmosphere Observations

For the present time, two fortuitous, related events have enabled observations of exoplanet atmospheres using a technique very different from direct imaging. The first event is the existence and discovery of a large population of planets orbiting very close to their host stars. These so-called hot Jupiters, hot Neptunes, and hot super Earths have up to about four-day orbits and semi-major axes less than 0.05 AU (see Figure 1). The hot Jupiters are heated by their parent stars to temperatures of 1000 to 2000 K, making their infrared brightness on the order of 1/1000 that of their parent stars (Figure 4).  While it is by no means an easy task to observe a 1:1000 planet-star flux contrast, such an observation is possible—and is unequivocally more favorable than the $10^{-10}$ visible-wavelength planet-star contrast for an Earth-twin orbiting a sun-like star.

The second favorable occurrence is that of transiting exoplanets—planets that pass in front of their star as seen from Earth. The closer the planet is to the parent star, the higher its probability to transit. Hence the existence of short-period planets has enabled the discovery of many transiting exoplanets. It is the special transit configuration that allows us to observe the planet atmosphere without imaging the planet.

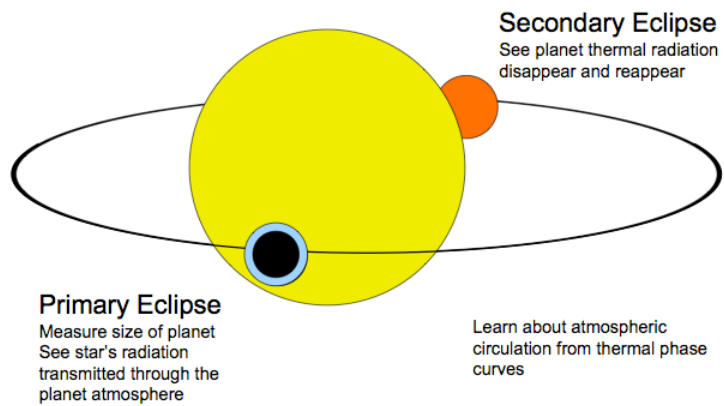

Transiting planets are observed in the combined light of the planet and star (Figure 5). As the planet passes in front of the star, the starlight drops by the amount of the planet-to-star area ratio. If the size of the star is known, the planet size can be determined. During transit, some of the starlight passes through the the planetary atmosphere (depicted by the annulus in Figure 5), picking up some of the spectral features in the planet atmosphere. A planetary transmission spectrum can be obtained by dividing the spectrum of the star and planet during transit by the spectrum of the star alone (the latter taken before or after transit).

Figure 5.  Schematic of a transiting exoplanet and potential follow-up measurements.  Note that primary eclipse is also called a transit.

Planets on circular orbits that pass in front of the star also disappear behind the star. Just before the planet goes behind the star, the planet and star can be observed together. When the planet disappears behind the star, the total flux from the planet-star system drops because the planet no longer contributes. The drop is related to both relative sizes of the planet and star and their relative brightnesses (at a given wavelength). The flux spectrum of the planet can be derived by subtracting the flux spectrum of the star alone (during secondary eclipse) from the flux spectrum of both the star and planet (just before and after secondary eclipse). The planet's flux gives information on the planetary atmospheric composition and temperature gradient (at infrared wavelengths) or albedo (at visible wavelengths).

Observations of transiting planets provide direct measurements of the planet by separating photons in time, rather than in space as does imaging (see Figure 5 and Figure 6). That is, observations are made of the planet and star together. (We do not favor the "combined light" terminology because ultimately the photons from the planet and star must be separated in some way. For transits and eclipses the photons are separated in time.) Primary and secondary eclipses enable high-contrast measurements because the precise on/off nature of the transit and secondary eclipse events provide an intrinsic calibration reference. This is one reason why the *HST* and the *Spitzer Space Telescope* (*Spitzer*) have been so successful in measuring high-contrast transit signals that were not considered in their designs.

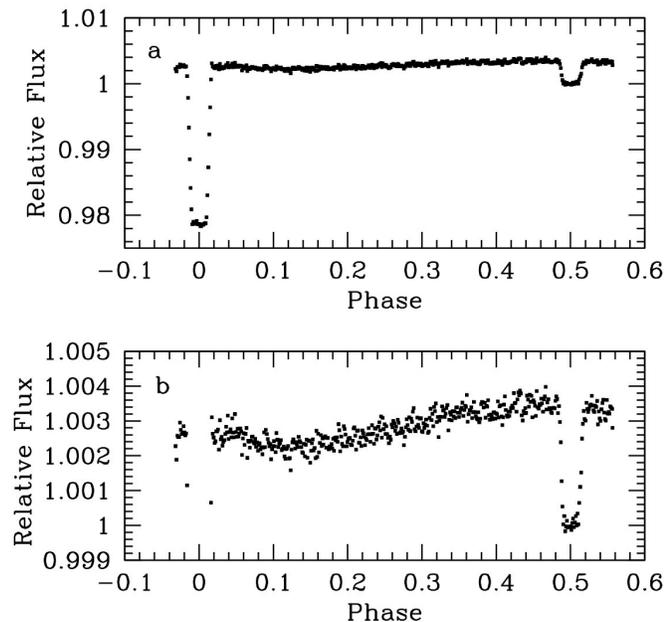

Figure 6 Infrared light curve of HD 189733A and b at 8 μm. The flux in this light curve is from the star and planet combined. Panel a: the first dip (from left to right) is the transit and the second dip is the secondary eclipse. Panel b: a zoom in of panel a. Error bars have been suppressed for clarity. Data from Knutson et al. (2007a).

## 2.2 Atmosphere Models and Theory

A range of models are used to predict and interpret exoplanet atmospheres. Usage of a hierarchy of models is always recommended. Interpreting observations and explaining simple physical phenomena with the most basic model that captures the relevant physics often lends the most support to an interpretation argument. More detailed and complex models can further support results from the more basic models. The material in this subsection is taken from Seager (2010).

***Computing a model spectrum.*** The equation of radiative transfer is the foundation not only to generating a theoretical spectrum but also to atmosphere theory and models. The radiative transfer equation is the change in a beam of intensity $dI/dz$ that is equal to losses from the beam $-\kappa I$ and gains to the beam $\varepsilon$, and the 1D plane-parallel form is

$$\mu \frac{dI(z,\nu,\mu,t)}{dz} = -\kappa(z,\nu,t)I(z,\nu,\mu,t) + \varepsilon(z,\nu,\mu,t).$$

Here: $I$ is the intensity [$Jm^{-2}s^{-1}Hz^{-1}sr^{-1}$], a beam of traveling photons; $\kappa$ is the absorption coefficient [$m^{-1}$] which includes both absorption and scattering out of the radiation beam; $\varepsilon$ is the emission coefficient [$Jm^{-3}s^{-1}Hz^{-1}sr^{-1}$] which includes emission and scattering into the beam; $\mu = \cos\theta$, where $\theta$ is the angle away from surface normal; and $z$ is vertical altitude, where each altitude layer has a specified temperature and pressure. Using the definition of optical depth $\tau$, $d\tau = -\kappa dz$, yields a common form of the radiative transfer equation,

$$\mu \frac{dI(\tau,\lambda,\mu,t)}{d\tau} = -\kappa(\tau,\lambda,t)I(\tau,\lambda,\mu,t) + B(\tau,\lambda,t).$$

Here we have omitted scattering, adopted LTE, and used Kirchoff's law $B = \varepsilon/\kappa$, to explicitly use the blackbody function $B$.

The solution of the radiative transfer equation has a long history in stellar and planetary atmosphere theory. Simplified solutions for exoplanet spectra are possible for: transmission spectra by the case of no emission (*i.e.*, $\varepsilon = 0$); and for atmospheric thermal emission spectra by ignoring scattering in the $\kappa$ and $\varepsilon$ terms. The significant differences in radiative transfer between exoplanet atmospheres and stellar atmospheres are the boundary condition at the top of the atmosphere, namely the incident stellar radiation, and the possibility of clouds for exoplanet atmospheres.

Inherent in the radiative transfer equation are opacity, chemistry, and clouds, via the absorption coefficient $\kappa$ and the emission coefficient $\varepsilon$. It is fair to say that almost all of the detailed physics and the unknowns are hidden in these macroscopic coefficients. The absorption coefficient for a single gas species is defined by $\kappa(\lambda, T, P) = n(T,P)\sigma(\lambda, T, P)$, where $n$ is the gas number density, and $\sigma(\lambda, T, P)$ is the cross section summed over all molecular lines that contribute at a given wavelength and that includes partition functions. At a given wavelength, $\kappa$ from all relevant gas molecules must be included. The cross sections themselves come from either laboratory measurements or from quantum mechanics calculations. For a description of opacity calculations relevant to exoplanets see Freedman, Marley, & Lodders (2008), Sharp & Burrows (2007), and the HITRAN database (Rothman et al. 2009).

The number density of a gas molecule can be calculated from equilibrium chemistry (e.g., Burrows & Sharp 1999; Lodders & Fegley 2002). In some cases, nonequilibrium chemistry is significant, especially for situations where the strong CO double bond and $N_2$ triple bond cannot be broken fast enough to reach chemical equilibrium (e.g., Saumon et al. 2006). Photochemistry is significant in driving the molecular abundances for low-mass rocky planets with thin atmospheres. Such planets cannot hold onto the light chemical species and do not have deep atmosphere temperatures and pressures to return photochemical products back to their equilibrium concentrations. Atmospheric escape is a further complication for the outcome of low-mass and/or hot exoplanet atmosphere composition and determines which elements remain in the atmosphere.

Clouds complicate the radiative transfer solution due to the often high opacity of solid material. The type of cloud that forms depends on the condensation temperature of the gas, and for hot Jupiters high-temperature condensates such as iron and silicates are likely present. For an excellent introduction to cloud physics see Sanchez-Lavega, Perez-Hoyos, & Hueso (2004).

Given an atmospheric temperature-pressure profile, a planet's emergent spectrum can be calculated with all of the above ingredients. In exoplanet atmospheres, a reality check comes from the idiom "what you put in is what you get out". This is a warning that arbitrary choices in inputs (molecular abundances and boundary conditions) can control the output spectrum.

***Computing a 1D temperature profile: radiative transfer, hydrostatic equilibrium and conservation of energy.*** In order to describe the temperature-pressure structure of a planetary atmosphere, three equations are needed. The equation of radiative transfer, the equation of hydrostatic equilibrium, and the equation of conservation of energy. With these three equations, three unknowns can be derived: temperature as a function of altitude; pressure as a function of altitude; and the radiation field as a function of altitude and wavelength. Hydrostatic equilibrium describes how the atmospheric pressure holds up the atmosphere against gravity, and relates pressure to altitude. Conservation of energy is described by radiative equilbrium in an altitude layer: energy is conserved because energy is neither created nor destroyed in the atmosphere. If the atmosphere layer(s) is unstable against convection, then convective equilibrium or radiative-convective equilibrium holds and is used to describe energy transportation in that layer. All of the chemistry, opacity, and cloud issues hold for the computation of the vertical temperature profile of a planet atmosphere because heating and cooling depends on the details of absorption and emission at different altitudes.

***Computing the 3D atmospheric structure: atmospheric circulation***
Atmospheric circulation is the large-scale movement of gas in a planetary atmosphere that is responsible for distributing energy absorbed from the star throughout the planetary atmosphere. The best current example where atmospheric circulation models are required are hot Jupiter exoplanets. With semi-major axes less than about 0.05 AU, hot Jupiters are expected to be tidally-locked to their host stars, having a permanent day and night side. Furthermore, because of the close stellar proximity, the planetary dayside is intensely irradiated by the host star, setting up a radiative forcing regime not seen in the solar system. An understanding of both the emergent spectra (e.g., Fortney et al. 2006) and the redistribution of absorbed stellar energy (e.g., Showman, Cho, & Menou 2010) require atmospheric circulation models.

Atmospheric circulation models are based on the fluid dynamic equations. They come from six fundamental equations: the conservation of mass, conservation of momentum (one equation for each dimension), conservation of energy, and the ideal gas law as the equation of state. Some investigators have used the full set of equations for exoplanet models. Others take the traditional planetary atmospheres approach resulting from four decades of study: the primitive equations. The primitive equations replace the vertical momentum equation with local hydrostatic balance, thereby dropping the vertical acceleration, advection, Coriolis, and metric terms that are generally expected to be less important for the global-scale circulation, such that energy is still conserved. Still other researchers use a 2D version, the shallow water equations. Because of the short timescales involved for radiation transport, atmospheric circulation models traditionally use very elementary radiation schemes that might not be suitable for hot Jupiters. This is recently changing as radiative transfer schemes are implemented (Showman et al. 2009; Menou & Rauscher 2009). The limiting factors of atmospheric circulation research are both the nonlinearity of the equations and the very computationally intensive models.

Atmospheric circulation is not always necessary to compute spectra for many types of planets (e.g., equatorially-viewed Earth, Jupiter), especially given the current quality of exoplanet data. Atmospheric circulation, however, is needed to understand planets with strong external radiative forcing. In addition, atmospheric circulation partly controls surface temperatures and drives large-scale cloud patterns (and hence albedo) on terrestrial planets. See Showman, Cho, & Menou (2010) for a description of all of these ideas as well as applications to both giant and terrestrial planets.

## 2.3 Anticipated Planet Atmosphere Diversity

The diversity of planet interior compositions is highly relevant to atmospheres, so we show in Figure 7 a generic summary of types of exoplanets via their bulk composition as a function of rock, ice, or gas components (Chambers 2010, Rogers & Seager 2009; see Seager et al. 2007 for a description of mass-radius relations for planet interior compositions).

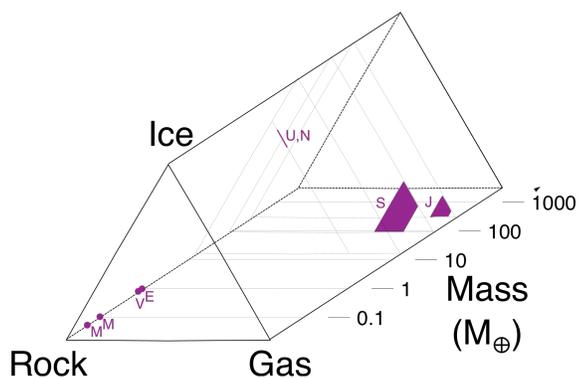

Figure 7. Schematic diagram illustrating the range of possible planet primordial bulk compositions for exoplanets. In this figure "gas" refers to primordial H and He accreted from the nebula, "ice" refers to ice- forming materials, and "rock" refers to refractory materials. Constraints on the current compositions of the solar system planets are plotted in purple (planets are denoted by their first initial). Exoplanets might appear anywhere in this diagram. Adapted from Chambers (2010) and Rogers and Seager (2009).

In the solar system there is a definite relationship between the relative abundances of rock-ice-gas and planet mass: small planets ($M \leq 1\,M_\oplus$) are rocky, intermediate planets ($\sim$15-17 $M_\oplus$) are icy, and larger planets are predominantly composed of H and He. Whether or not exoplanets also follow this pattern is one of the most significant questions of exoplanet formation, migration and evolution.

Exoplanet atmospheres are related to their interiors, but how much so remains an outstanding question. The best way to categorize exoplanet atmospheres in advance of detailed observations is within a framework of atmospheric content based on the presence or absence of volatiles (see Rogers and Seager for a discussion focusing on a specific exoplanet, GJ 1214b). We choose five categories of atmospheres:

1. Dominated by H and He. Planetary atmospheres that predominantly contain both H and He, in approximately cosmic proportions are atmospheres indicative of capture from the protoplanetary nebula (or planet formation from gravitational collapse). In our solar system these include the giant and ice giant planets

2. Outgassed atmospheres with H. Planets that have atmospheres from outgassing and not captured from the nebular disk will have some hydrogen content in the form of $H_2$. How much depends on the composition of planetesimals from which the planet formed (Elkins-Tanton & Seager 2008; Schaefer and Fegley 2010). The idea is that some planets in the mass range 10 to 30 Earth masses will be massive enough and cold enough to retain hydrogen in their atmospheres against atmospheric escape. Such H-rich atmospheres will have a different set of dominating molecules ($H_2$, naturally occurring $H_2O$, and $CH_4$ or CO) as compared to solar system terrestrial planets with $CO_2$ or $N_2$ dominated atmospheres. Some super Earths may have outgassed thick atmospheres of up to 50% by mass of H, up to a few percent of the planet mass. Other planets may have massive water vapor atmospheres (e.g., Leger et al. 2004; Rogers & Seager 2009). Outgassed atmospheres will not have He, since He is not trapped in rocks and cannot be accreted during terrestrial planet formation (Elkins-Tanton and Seager 2008).

3. Outgassed atmospheres dominated by $CO_2$. On Earth $CO_2$ dissolved in the ocean and became sequestered in limestone sedimentary rocks, leaving $N_2$ as the dominant atmospheric gas. This third category of atmospheres, stemming from either the first or second category, is populated by atmospheres that have lost hydrogen and helium, wherein signs of $H_2O$ will be indicative of a liquid water ocean. The actual planet atmospheric composition via outgassing depends on the interior composition.

4. Hot super Earth atmospheres lacking volatiles. With atmospheric temperatures well over 1500 K, hot Earths or super Earths will have lost not only hydrogen but also other volatiles such as C, N, O, S. The atmosphere would then be composed of silicates enriched in more refractory elements such as Ca, Al, Ti (Schaefer & Fegley 2009).

5. Atmosphereless planets. A fifth and final category are hot planets that have lost their atmospheres entirely. Such planets may have a negligible exosphere, like Mercury and the Moon. Transiting planets lacking atmospheres can be identified by a substellar point hotspot (e.g., Seager & Deming 2009); for planets with atmospheres the hot spot is likely to be advected from the substellar point and for planets with thick atmospheres the planet might not be tidally locked.

The actual atmospheric molecular details of the above scenarios must await theoretical calculations that include outgassing models, photochemistry calculations, atmospheric escape computations, and also future observations. Understanding enough detail to create such a classification scheme may well occupy the field of exoplanet atmospheres for decades to come.

At the present time there is a great divide between the hot Jupiter exoplanets that we can study observationally and the super Earths that we want to study but which are not yet accessible. We begin with observation and interpretation highlights of hot Jupiters.

# 3. Discovery Highlights

We now turn to a summary of the most significant exoplanet atmospheric discoveries. Hot Jupiters dominate exoplanet atmosphere science, because their large radii and extended atmospheric scale heights facilitate atmospheric measurements to maximum signal-to-noise. Even so, data for hot Jupiters remain limited, so their physical picture cannot yet be certainly described. We here address what we have learned from the observations alone, as well as from interpreting the observations with the help of models. In so doing, we rely on the formal uncertainties of each observation as reported in the literature. We discuss these conclusions starting with the most robust, and working toward more tentative results.

## 3.1. Hot Jupiters are Hot and Dark

Hot Jupiters are blasted with radiation from the host star. They should therefore be kinetically hot, heated externally by the stellar irradiance. Indeed, early hot Jupiter model atmospheres already predicted temperatures exceeding 1000K (Seager & Sasselov 1998; Sudarsky, Burrows, & Hubeny 2003). The first and most basic conclusion from the *Spitzer* secondary eclipse detections (e.g., Figure 6) was the confirmation of this basic paradigm. The fact that the planets emit generously in the infrared implies that they efficiently absorb visible light from their stars. Searches for the reflected component of their energy budget have indicated that the planets must be very dark in visible light, with geometric albedos less than about 0.2 (Rowe et al. 2008), and likely much lower. Purely gaseous atmospheres lacking reflective clouds can be very dark (Marley et al. 1999; Seager, Whitney, & Sasselov 2000) but HD209458b also requires a high-altitude absorbing layer (see below) to account for its atmospheric temperature structure.

### 3.2 Identification of Atoms and Molecules

A planetary atmosphere with elemental composition close to solar and heated upwards of 1000 K is expected to be dominated by the molecules $H_2$, $H_2O$, and, depending on the temperature and metallicity, CO and/or $CH_4$. Of these molecules, $H_2O$ is by far the most spectroscopically active gas. Water vapor is therefore expected to be the most significant spectral feature in a hot Jupiter atmosphere. Some initial indications from *Spitzer* spectroscopy that water absorption was absent (Richardson et al. 2007; Grillmair et al. 2007) were superceded by higher S/N data that clearly showed water absorption (Grillmair et al. 2008; Swain et al. 2008, Swain et al 2009a). See Figures 8 and 9.

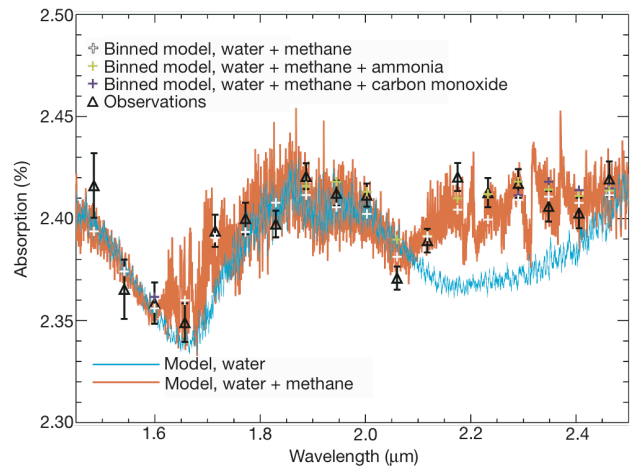

Figure 8. Transmission spectrum of the transiting planet HD 189733. *Hubble Space Telescope* observations shown by the black triangles. Two different models highlight the presence of methane in the planetary atmosphere. From Swain et al. (2008).

Other atoms and molecules identified in hot Jupiter atmospheres are atomic sodium (Na) (e.g., Charbonneau et al. 2002; Redfield et al. 2008), methane ($CH_4$) (Swain et al. 2008), carbon monoxide (CO), and carbon dioxide ($CO_2$) (e.g., Swain et al. 2009a,b; Madhusudhan and Seager 2009). This set of molecules has been detected in the two hot Jupiters most favorable for observation (HD 209458b and HD 189733b). It is instructive to consider which atomic and molecular identifications are model-independent and which depend on models. The atomic sodium detections are independent of models, because there is no other plausible absorber at the sodium-doublet wavelength (see the analysis in Charbonneau et al. 2002). The *HST* spectrophotometry has sufficiently high spectral resolution to show distinct features from $H_2O$, $CH_4$, and $CO_2$. Those detections may also be considered model-independent because they rely only on molecular absorption cross section information.

Planets with host stars fainter than approximately $V$=8 have been observed primarily using broad-band photometry, since they do not produce enough photons to be observed to the requisite precision with *HST*/NICMOS or the *Spitzer*/IRS instrument. From broad-band photometry, not only are model fits needed to identify molecules, but some assumptions as to which spectroscopically active molecules are present is essential. Taking $H_2O$, $CH_4$, CO and $CO_2$ as the molecules with features in the *Spitzer* 3.5-24 $\mu$m range, molecular identification is possible for $H_2O$ and $CH_4$ from the four

*Spitzer*/IRAC bandpasses between 3.4 and 8 $\mu$m and for $H_2O$, $CH_4$, CO and $CO_2$ with the additional *Spitzer*/IRS 16 $\mu$m *Spitzer*/MIPS 24 $\mu$m bandpass (see Figure 9).

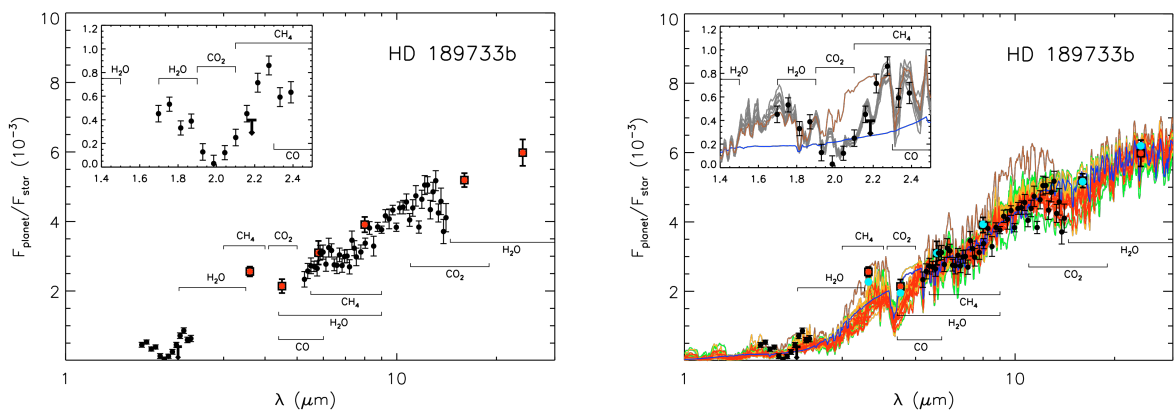

Figure 9. Thermal emission data composite for HD 189733 in secondary eclipse. Data from *HST*/NICMOS (inset: Swain et al. 2009), *Spitzer*/IRAC (four shortest wavelength red points; Charbonneau et al. 2008), Spitzer/IRS-PU (Deming et al. 2006), *Spitzer*/MIPS (Charbonneau et al. 2008), *Spitzer*/IRS (black points from 5 – 13 microns; Grillmair et al. 2008). Models shown in the bottom panel (from Madhusudhan and Seager 2009) illustrate that the best fits to the *Spitzer*/IRS ((red curve shows fits within the 1.4σ errors, on average; orange 1.7σ, green 2σ, and blue is one best fit model within 1.4σ) and *Spitzer* photometry (brown curve within 1σ) do not fit the NICMOS data (inset grey curves within 1.4σ) possibly implying variability in the planet atmosphere from data taken at different epochs. For abundance constraints from the difference models, see Madhusudan and Seager (2009).

A thorough temperature and abundance retrieval method enables statistical constraints on molecular mixing ratios and other atmospheric properties (e.g., Madhusudhan and Seager, 2009). As a best case example, the *HST*/NICMOS spectrum of 189733b at secondary eclipse yields constraints of $H_2O \sim 10^{-4}$, considering fits within the ~ 1.5 σ observational uncertainties. For other species and other data sets, the constraints are not nearly as good. At the same level of fit, the six-channel Spitzer photometry of HD 209458b at secondary eclipse yields constraints of $H_2O < 10^{-4}$, $CH_4 > 10^{-8}$, $CO > 4 \times 10^{-5}$, and $2 \times 10^{-9} < CO_2 < 7 \times 10^{-6}$. The constraints placed to date have been limited by the number of simultaneous broadband observations available, which are typically fewer than the number of model parameters.

The inference of $CO_2$ is interesting, because a planetary atmosphere dominated by molecular hydrogen is expected to have CO or $CH_4$ as the primary reservoir of carbon at high temperatures. For a hot atmosphere with CO as the primary carbon reservoir, $CO_2$ could be reasonably abundant ~$10^{-6}$, based on thermochemistry (Lodders & Fegley 2002) or photochemistry (Liang et al. 2003), but higher values require a high planet metallicity (Zahnle et al. 2009). $CO_2$ mixing ratios of $10^{-4}$ or higher might be needed to explain the observed $CO_2$ features in the *HST*/NICMOS dataset for the HD 189733 thermal emission spectrum (Madhusudhan and Seager 2009; c.f. Swain et al. 2009a).

In addition to molecules, the presence of atmospheric haze has been inferred in HD 189733 via transmission spectra with HST/STIS. While the particle composition has not been identified, the Rayleigh-scattering behavior of the data indicates small particle sizes (Pont et al. 2008). The presence of haze on HD189733b is consistent with similar inferences for HD209458b. In the latter planet, high clouds or haze have been invoked to account for the weakness of the sodium absorption (Charbonneau et al. 2002), and the upper limits on CO absorption (Deming et al. 2005b), during transit.

### 3.3 Day-Night Temperature Gradients

Hot Jupiters are fascinating fluid dynamics laboratories because they probably have a permanent dayside and a permanent night side. Close-in giant planets are theorized to have their rotation synchronized with their orbital motion by tidal forces, a process that should conclude within millions of years (e.g., Guillot et al. 1996). Under this tidal-locking condition the planet will keep one hemisphere perpetually pointed toward the star, with the opposite hemisphere perpetually in darkness.

A resulting key question about tidally-locked hot Jupiters concerns whether one side of the planet is extremely hot, and the other side remains very cold. Or does atmospheric circulation even out the planetary day to night side temperature difference? Observational evidence exists for planets approaching both extremes. *Spitzer* thermal infrared observations of HD 189733b show that the planet has only a moderate temperature variation from the day to night side. The planet shows an 8 $\mu$m brightness temperature variation of over 200 K from a minimum brightness temperature of 973±33K to a maximum brightness temperature of 1212±11K (Knutson et al. 2007a), and a thermal brightness change at 24 $\mu$m consistent with the 8 $\mu$m data within the errors (Knutson et al. 2009c). In contrast to HD 189733b, Ups And and HAT-P-7 show a dramatic change in thermal brightness from the day to the night side. Although the Ups And data are not continuous, the brightness change at 8 $\mu$m measured with *Spitzer* indicates a temperature change of well over 1000 K. HAT-P-7b, with a dayside equilibrium temperature above 2,000 K, has a spectacular phase curve measured by the *Kepler Space Telescope* (Borucki et al. 2010). The HAT-P-7b data are difficult to interpret in terms of day-night temperature gradient because *Kepler*'s single broad bandpass (from 420 to 900 nm) allows both thermal emission and reflected radiation to contribute to the observed signal, and because visible-wavelength thermal emission at the Wien tail changes rapidly for a ~2,000 K black body radiator decreasing in temperature.

A word of caution is warranted concerning the day-night temperature gradient inferred from broad-band photometric light curves. One could imagine a malevolent situation where an emission band present on the planetary dayside turns into an absorption feature on the planetary night side. This scenario would mimic a large *horizontal* thermal gradient but would actually be caused by a variation in vertical temperature gradient, without the need for significant day-night temperature differences.

One particularly interesting result that is relevant to day-night temperature differences is *Spitzer*'s observation of the periastron passage of the very eccentric (*e*=0.93) planet HD80606b (Laughlin et al. 2009). Radiative cooling plays a role in day-night temperature differences, and Spitzer's observed time-dependent heating of this planet at periastron constrains the atmospheric radiative time constant. Moreover, because rotation modulates the flux we receive from the sub-stellar hemisphere at periastron, those observations also provide key information on the planet's pseudo-synchronous rotation. The pseudo-synchronous rotation rate in turn is sensitive to the physics of energy dissipation in the planetary interior. So *Spitzer* observations of eccentric planets can in principle provide information on planetary interiors as well as atmospheres.

Turning to model interpretation of the planet HD 189733b, *Spitzer* thermal emission light curve observations indicate that strong winds have advected the hottest region to the east of the sub-stellar point (Knutson et al. 2007a; Showman et al. 2009). The shifted hot region on the dayside carries physical information such as the speed of the zonal circulation, and information about the altitude and opacity-dependence of the atmospheric radiative time constant.

Theory alone can articulate a few significant points about atmospheric circulation. Giant planets in our solar system also have strong zonal winds, appearing in multiple bands at different latitudes. Consideration of the relatively slow rotation rate of hot Jupiters (probably equal to their orbital period of a few days) leads us to believe that their zonal winds will occur predominantly in one or two major jets that are quite extended in latitude and longitude (e.g., Showman & Guillot 2002; Menou et al. 2003). The relatively large spatial scale of hot Jupiter winds—and the corresponding temperature field— should be a boon to their observational characterization. Planets with temperature fluctuations only on small scales will have little to no variation in the amount of their hemisphere-averaged flux as a function of orbital phase. Atmospheric circulation models also show that it seems likely that at least some hot Jupiters transport energy horizontally via zonal winds having speeds comparable to the speed of sound (e.g., Showman et al. 2010).

## 3.4 Atmospheric Escape

Escaping atomic hydrogen from the exosphere of the hot Jupiter HD 209458b has been detected during transit in the Ly $\alpha$ line. A positive detection was made with *HST*/STIS (3.75 $\sigma$) *HST*/STIS (Vidal-Madjar et al. 2003). (Other absorption features claimed at lower statistical significance are not discussed here.) Showing a 15% drop in stellar Ly $\alpha$ intensity during transit, the HD 209458b observations are interpreted as a large cloud of hot hydrogen surrounding the planet. The cloud extends up to 4 planetary radii, and the kinetic temperature is as high as tens of thousands of K. It extends beyond the planetary Roche lobe, so the hydrogen is evidently escaping to form a possible comet-like coma surrounding the planet. Models agree that the implied exospheric heating is likely due to absorption of UV stellar flux, but Jeans escape is not sufficient to account for the hydrogen cloud. The specific origin of the escaping atoms is model-dependent. Escape mechanisms include radiation pressure, charge exchange, and solar wind interaction (see, e.g. Lammer et al. 2009 and references therein).

### 3.5 Vertical Thermal Inversions

Almost all solar system planets have thermal inversions high in their atmospheres (i.e., the temperature is rising with height above the surface). These so-called stratospheres are due to absorption of UV solar radiation by $CH_4$ -induced hazes or $O_3$. Thermal inversions in hot Jupiter atmospheres were not widely predicted, because of the expected absence of $CH_4$, hydrocarbon hazes, and $O_3$.

Evidence for vertical atmospheric thermal inversions in hot Jupiters comes from emission features in place of (or together with) absorption features in the thermal infrared spectrum (for a basic explanation see Seager 2010). Because broad-band photometry does not delineate the structure of molecular spectral bands, the inference of a thermal inversion must rely on models. Moreover, the presence of specific molecules must be inferred based on the anticipated physical conditions, guided by models. *Spitzer* data show that the upper atmospheres of several planets have thermal inversions, if water vapor is present and if abundances are close to solar (e.g., Burrows et al. 2008; see Figure 10).

The hot Jupiter temperature inversions are likely fueled by absorption of stellar irradiance in a high-altitude absorbing layer. Possibilities for a high-altitude absorber include gaseous TiO and VO (Hubeny, Burrows, & Sudarsky 2003; Fortney et al. 2007), as well as possibilities involving photochemical hazes (Zahnle et al. 2009). Under a simple irradiation-driven scenario, the stronger the stellar irradiance, the more likely that an inversion would occur (Hubeny, Burrows, & Sudarsky 2003; Fortney et al. 2007; Burrows, Hubeny, Budaj 2008). In addition, under this irradiation-driven scenario, planets with strong thermal inversions are also expected to show strong day-night temperature gradients. Hot Jupiters are probably more complex than this simple division allows; HD 189733b and XO-1b have virtually identical levels of irradiation and yet XO-1b appears to have an inversion (Machalek et al. 2008) while HD 189733b does not (Charbonneau et al. 2008). So it is possible that not-yet-understood chemistry may be a more dominant factor than stellar irradiance.

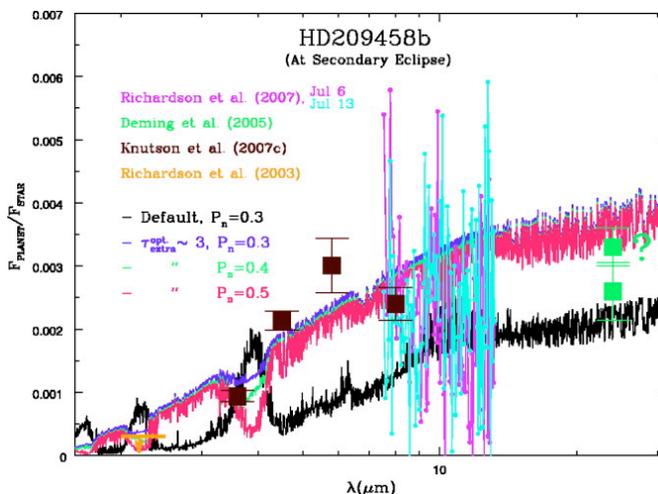

Figure 10. Evidence for an atmospheric thermal inversion for HD 209458b. *Spitzer* data points from secondary eclipse measurements are shown with brown (IRAC; Knutson et al. 2007) and green (Deming et al. 2005 and private comm.; the two points are data taken at different times). IRS spectra shown in purple and aqua are from Richardson et al. (2007). The model in pink shows emission features from an atmospheric thermal inversion. The black curve is a non-thermal-inversion model. Figure from Burrows et al. (2007).

Taking a critical look at the evidence for thermal inversions, (Madhusudhan & Seager (2010)) found that for many cases existing observations (*Spitzer* broad-band photometry) are not enough to make robust claims on the presence of thermal inversions. For some of these cases, the observations can be explained without thermal inversions, along with rather plausible chemical compositions. For other cases, the observations can be explained by models without thermal inversions if the models have a high abundance of $CH_4$ instead of CO. The dominance of methane as the carbon-bearing molecule would indicate severe non-equilibrium chemistry, since at high temperatures CO is the most stable, and hence most abundant, carbon-bearing molecule. 3D atmospheric circulation models also have difficulty explaining the *Spitzer* IRAC data of HD 209458b and other planets with purported thermal inversions (Showman et al. 2009). The models produce day-night redistribution over a continuous range of pressures; as a result the circulation models produce temperature profiles that do not explain the IRAC data, even if a thermal inversion is present at some locations (Showman et al. 2009).

Determination of thermal inversions will be more solid and less model-dependent and degenerate when higher spectral resolution and higher S/N observations become available using *JWST*, or possibly advanced ground-based facilities.

### 3.6 Variability

Variability in the hot Jupiter atmospheric data is relatively common at the $2\sigma$ level. While this would not be statistically significant in a particular case, it is highly suggestive in the aggregate, because it occurs more frequently than is expected based on statistical error distributions. A possible mundane explanation is that the observers have underestimated their statistical errors. Nevertheless, many observers use careful methods for error estimation, so we must consider the intriguing possibility that the atmospheres of hot Jupiters are intrinsically variable to a significant degree. Moreover, there are some specific examples where variability seems inescapable, unless there are major systematic errors in the observations. For example, the $CO_2$ feature detected in HD189733b at 2.3 $\mu$m with HST/NICMOS during both transit (Swain et al. 2008) and eclipse (Swain et al. 2009) is far too strong to be consistent with the lack of an observed $CO_2$ absorption at 4 $\mu$m and 16 $\mu$m in secondary eclipse observations (Charbonneau et al. 2008). The $CO_2$ absorption cross sections are much greater at 4- and 16- than at 2.3 $\mu$m, so this conclusion is not significantly model-dependent (see Madhusudhan and Seager 2009).

# 4 Observational Challenges for Transiting Planet Observations

Hand-in-hand with the exoplanet atmosphere discovery highlights based on observations are issues of systematic effect removal. The *HST* and *Spitzer* were not designed to achieve the very high signal-to-noise ratios necessary to study transiting exoplanets, but they have succeeded nonetheless. We have learned from these observations that with enough photons, systematics that were unknown in advance can often be corrected, as long as the time scale and nature of the systematic effects do not overlap too strongly with the time scale of the signals being sought. A cautionary view is that observers are pushing the limit of telescopes and instruments far past their design specifications. The art and science of exoplanet atmospheric observations, therefore, is instrument systematic removal to extract a planetary atmospheric signal. Extreme care must be taken to reach realistic results and to assign appropriate error limits.

## 4.1 Systematic Observational Errors and Their Removal

Separating the light of exoplanets from that of their star is difficult. The contrast ratio (planet divided by star) is greater in the infrared than in the optical, but it can still be as small as $10^{-4}$, or even less (Figure 4). Because transit and eclipse methods reply on separating the planetary and stellar light in the time domain, we have to be concerned with the temporal stability of the measurements at the $10^{-4}$ level or less. Also, astronomers want to push the measurements to increasingly smaller planets, that have even lower planet-star contrast ratios than the currently-studied hot Jupiters. Inevitably, then, we are going to have to deal with systematic observational errors caused by temporal instabilities. Even for hot, giant exoplanets, these systematic errors are already a significant issue.

Systematic errors are not random, they are signals in their own right. The key to their removal is in understanding the origin and nature of those signals, so that they can be modeled and removed from the data, leaving the planetary signal undistorted by the process. Nearly all of the facilities that have measured light from exoplanets are general purpose: *HST*, *Spitzer*, and ground-based observatories. NASA's *Kepler Space Telescope* is an exception to the general purpose telescope in that *Kepler* was specifically designed to acquire stable, precise stellar photometry. See Caldwell et al. (2010) for a report on *Kepler* systematics. Most other facilities, however, have some sources of systematic error in common, such as pointing instabilities, but the details differ with the specific telescope/instrument/detector and observing methodology.

### 4.1.1 *Spitzer*

*Spitzer* is among the most widely-used facilities for exoplanet characterization and so its systematic effects make particularly good examples. Overall, *Spitzer* is remarkably stable due to its cryogenic state and its placement in the thermally stable environment of heliocentric orbit. Also minimizing *Spitzer* systematics is the fact that the *Spitzer* instruments have very few moving parts. *Spitzer* results have therefore overall been robust, with uncorrected systematic errors at relatively low levels.

The bulk of *Spitzer* exoplanet work has used the IRAC instrument (Fazio et al. 2005). *Spitzer*/IRAC has four photometric bands: 3.6, 4.5, 5.8, and 8.0 μm. The 3.6 and 4.5 μm bands use InSb detectors, and the 5.8 and 8.0 μm bands use Si:As blocked-impurity-band detectors. The *Spitzer*/IRAC InSb detectors have significant intra-pixel sensitivity variations, wherein the sensitivity of each pixel is greatest at pixel center, and declines toward the edges of the pixel, typically by several percent (Morales-Calderon et al. 2006). Exoplanet investigations have been among the most useful in defining the nature of this systematic error. We have learned that the intrapixel variation is not symmetric about the center of each pixel; sensitivity gradients are larger in one direction than in the orthogonal direction. Moreover, the amplitude of the effect varies from pixel-to-pixel. The telescope pointing has a small amount of jitter on an approximately one hour time scale, and that causes the star's position on the pixel grid to vary by a significant fraction of a pixel. Moving the star on a pixel with spatial gradients in sensitivity leads to a time-variable signal level that each exoplanet investigation has to model and remove (e.g., Figure 11 and Desert et al. 2009).

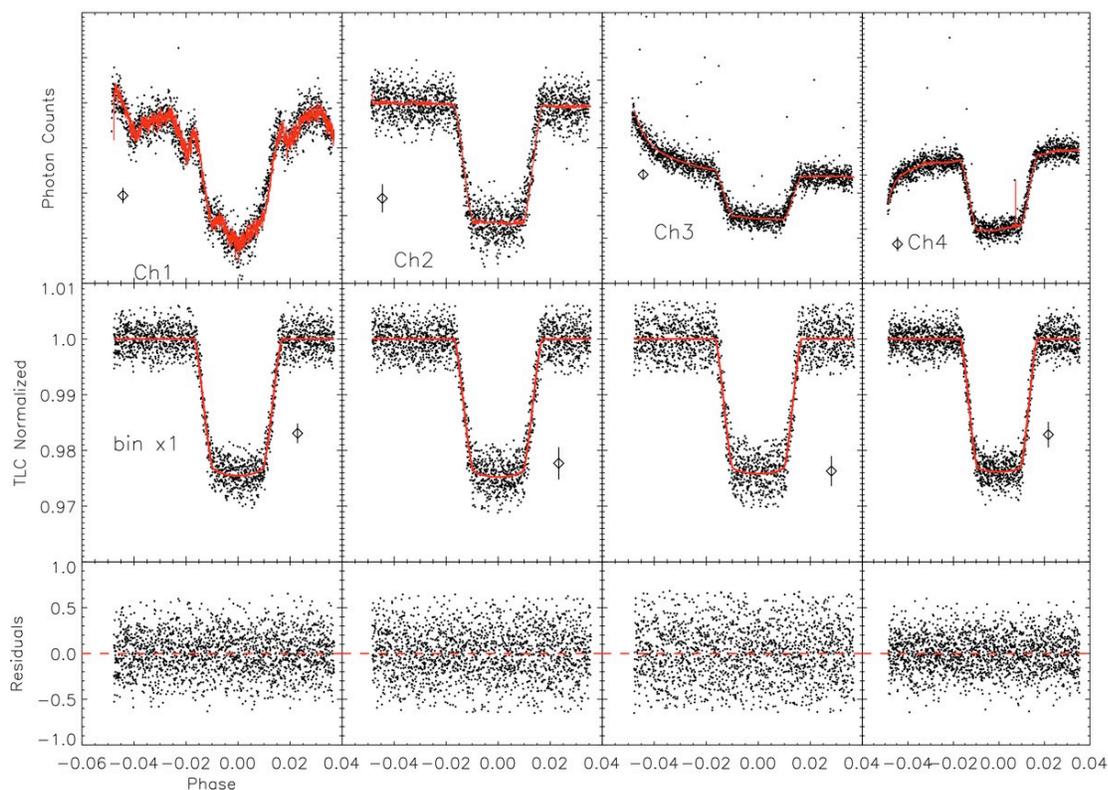

Figure 11. Illustration of *Spitzer*/IRAC systematic effects. Transit light curves, fits, and residuals for each channel. Data are not rebinned. Typical 1σ error bars on individual measurements are represented. The raw-weighted light curves (top panel) have to be corrected for large fluctuations correlated to the "pixel phase" and to the ramp baseline. The corrected light curve is plotted in the middle panel without the rejected points (rejection at more than 3σ). Overplotted is the fit with limb darkening taken into account. The bottom plot shows the residuals from the best fits. Channel 1 (3.5 μm) shows the intrapixel effect, channel 2 (4.6 μm) for this particular observation is fortuitously is a "magic pixel" immune from any intrapixel effects, channel 3 (5 μm) shows the "inverse ramp", and channel 4 (8 μm) shows the ramp which is associated with charge trapping. From Desert et al. (2009).

The removal is accomplished by decorrelating measured flux with either the X or Y coordinate, or the radial distance from pixel center. There is no universally accepted method for performing these decorrelations. For example, some investigators use radial distance from pixel center as the independent decorrelation variable, while other investigators use both X and Y position separately. Each investigator selects the decorrelation method that minimizes the noise in their own data. Since no two *Spitzer* observations place stars at the exact same position on the detector, and the properties of the detector vary from pixel to pixel, it is not surprising that the decorrelation methodology differs from one investigation to another. Whether this intrapixel source of systematic error is adequately removed depends on factors such as how well the spatial positions on the detector are sampled in any given dataset, and how much data are available to accomplish full sampling.

In the cryogenic mission, *Spitzer*'s intra-pixel sensitivity effects were usually modeled and removed based on the actual science data for each transit and/or eclipse. The telescope pointing jitter interacts with the intra-pixel sensitivity variations to modulate the observed stellar intensity, and this process is usually adequately sampled during a single eclipse observation. During *Spitzer*'s non-cryogenic phase, several investigations plan to "map the pixel" by deliberately rastering the star in a specific pattern to define the intra-pixel effect more completely, independently of specific observations. This should allow us to verify that the 3.6 and 4.5 $\mu$m decorrelation methods used to date have been sound.

The most prominent and well-studied of *Spitzer*'s systematic errors is the baseline effect called "the ramp" (e.g., Deming et al. 2006; Figures 11 and 12). This effect is a property of the Si:As detectors, and is most prominent in the 8.0 micron band. An example of the ramp is shown in Figure 12, with a transit and eclipse of the exoplanet GJ 436b. The ramp behaves as if the sensitivity of the detector is increasing with time, reaching a plateau after many hours. A transit or eclipse curve is superposed on this increasing baseline. There is no danger that the ramp could cause a false-positive detection of a weak eclipse, because its shape does not resemble an eclipse in any way. However, the ramp makes it more problematic to define the out-of-eclipse (or transit) level, and thus adds uncertainty to the eclipse (or transit) depth. The ramp becomes particularly important for weak eclipses, where a small error in the baseline can produce a relatively large fractional error in the fitted eclipse amplitude.

The steepness of the ramp depends on the prior illumination history of the detector: when a bright source was observed previously, the ramp is relatively flat. When the detector was previously exposed only to faint sources, or blank sky, the ramp is steeper. The ramp rises faster, and reaches a plateau level faster, at brighter levels of illumination. These properties led to the concept (Knutson et al. 2007a) that the ramp is due to charge trapping in the detector. The first electrons generated by incoming photons are captured by ionized impurities—the charge traps, so those electrons do not result in signal output. As photons flood the detector, these charge traps become full, and the detector achieves full sensitivity.

The charge trapping picture has been a useful concept in understanding the ramp, but there are several problems. First, the ramp from the same type of detector used at 5.8 μm is *negative*, i.e., the signal *decreases* with time (and that is sometimes also true at 8 μm). That change in behavior is difficult to fit into a simple charge-trapping concept. Second, we do not have an understanding of the ramp based on the solid-state (quantum) physics of the detector. The physics of the detector is extremely complex because the detector has multiple layers, and photons of different wavelength penetrate to different depths. Moreover, small changes in the unmeasured composition of the layers can have large effects. Consequently, we may never be able to understand the ramp from first principles. Second, there is no 100 percent satisfactory mathematical formula that reproduces the shape of the ramp accurately, and how that shape changes with illumination level. Log or exponential functions are usually used, and they given generally good fits (but c.f. Section 4.2.1). But such fits are clearly not rigorously correct since (among other reasons) they imply unphysical detector behavior—such as monotonically increasing detector sensitivity as time increases without bound. Third, because different researchers use different functions to fit the observed ramp, small levels of systematic error in deep eclipses or transits, and larger relative errors in weak eclipses or transits, are possible. This situation could be improved by finding a mathematical model for the ramp shapes that successfully fits them as a function of illumination level.

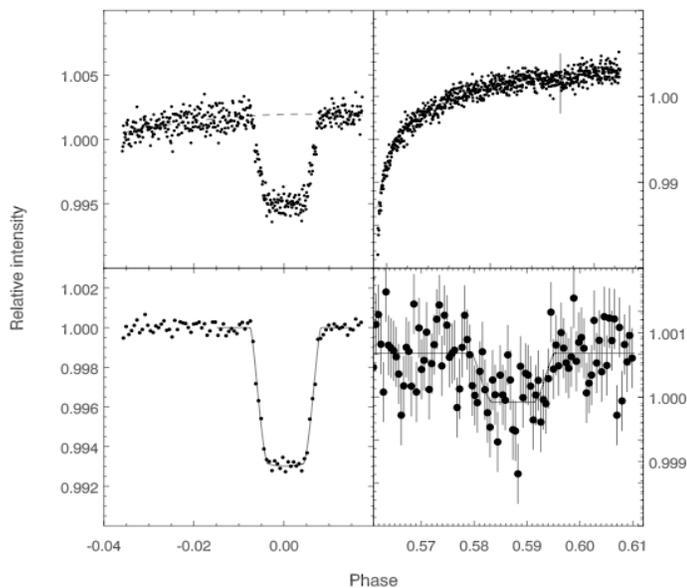

Figure 12. Illustration of the *Spitzer*/IRAC 8 μm ramp effect related to the pre-program observations. Shown are observations of the transit (left panels) and eclipse (right panels) of the hot Neptune planet GJ436b (Deming et al. 2007). The lower panel in each case shows the transit or eclipse after fitting and removing the ramp function. Note that the slope of the ramp is relatively flat for the transit observations, attributed to the fact that a bright source was observed just prior to these GJ436 observations. The steeper ramp for the eclipse is the normal situation for Spitzer observations at 8 microns.

A method to mitigate the ramp is to "preflash" the detector before an observation by observing a bright extended region such as a compact HII region. This technique exploits the observed property of the ramp that it is often flatter if the previous source was bright.  A bright source floods the detector with high flux levels and thus forces the ramp to its asymptotic value prior to science observations. (see, e.g., Knutson et al. 2009, Seager and Deming 2009).

### 4.1.2 The *Hubble Space Telescope*

*HST* has more complex systematic effects than *Spitzer*.  In contrast to *Spitzer*'s Earth-trailing heliocentric orbit, *HST*'s low-Earth orbit and 90 minute day/night orbit places it in a thermally variable environment. Unlike *Spitzer*'s instruments that have few moving parts, *HST*'s instrumental mechanical effects contribute to systematic errors, *e.g.* the cycling of the NICMOS filter wheel during Earth orbit can lead to errors at the quarter to half a percent level (e.g., Swain et al. 2008; Carter et al. 2009).

Exoplanet investigators have found that the first orbit in a new pointing usually has particularly high levels of systematic drifts in spectral and photometric parameters that interfere with exoplanet signals. Many of the systematic effects are highly correlated signals—correlated on timescales of the *HST* orbit. Decorrelating *HST*'s systematic effects requires using multiple independent variables.  For example, Pont et al. (2008) used multi-variable linear regression to remove systematic effects in ACS grism spectra that were a function of six independent variables: X- and Y-position of the spectrum, width and rotation of the spectrum, telescope orbital phase, and time.

There is not a complete attribution for the source of all *HST* systematics, although some are known to be caused by thermal variability and some specific described instrumental effects; this is in contrast to the *Spitzer* identified systematics of intrapixel variation and charge trapping. Because of the lack of association with physical processes and the complexity of the *HST* decorrelation techniques, it is important to cross-check *HST* results by using different instruments and methodologies.  So far, a successful cross-check using different instruments has not been accomplished. See Swain et al. (2008, 2009a, 2009b), Carter et al. (2009), and Pont et al. (2009) for a detailed description on *HST*/NICMOS systematics and their removal.

### 4.1.3 Ground-Based Observations

The Earth's atmosphere is, not surprisingly, the principal source of systematic error in ground-based exoplanet data.  These error sources differ depending on the technique used (i.e., spectroscopy or photometry), but some prominent sources include scintillation noise, uncorrelated angular and temporal fluctuations in atmospheric transparency, PSF changes due to variable seeing, and wavelength-dependent variations in seeing.  However, investigators pursuing accurate transit photometry (e.g., Winn et al., 2009) have achieved very high levels of precision, with scatter that averages down as the square-root of the number of samples—a sign that systematic error levels are low.  This success has been achieved by performing photometry relative to multiple comparison stars having a brightness comparable to the planet-hosting star, and mitigating the effects of seeing by using a moderate telescope defocus.

## 4.2 Discrepancies in the Literature

Confusion has grown as to which atmosphere detections are robust, due to a few conflicting reports in the literature. In order to help clarify this issue, we review three major discrepancies reported in the literature. One discrepancy involves a difference in removal of systematic effects in *Spitzer* data between different teams of investigators. Another involves possible systematic errors in *HST* data. An additional discrepancy is a refutation of a ground-based polarization detection. We also describe a set of observations not disputed in the literature but not widely accepted in the community, having to do with underestimated errors at the edges of *HST*/STIS spectral orders and related spectral "features".

### 4.2.1 *Spitzer* Infrared Photometry

*Spitzer* infrared transit photometry has given conflicting results for the effect of water absorption during transit for the planet HD 189733b. One group (Tinetti et al. 2007 and Beaulieu et al. 2008) finds a deeper transit at 5.8 $\mu$m than another group (Desert et al. 2009). A deeper 5.8 $\mu$m transit relative to other *Spitzer* bands would indicate that the atmosphere contains water to a sufficiently high altitude to increase the projected size of the optically thick atmospheric annulus at wavelengths like 5.8 $\mu$m where water vapor absorbs strongly. The difference in conclusions are most likely related to the way in which each group models and removes *Spitzer*'s systematic effects, specifically the functional form used for ramp removal, and whether the steepest early part of the ramp is a help, or a hindrance, to the best solution. See Beaulieu et al. (2010) and Desert et al. (2009) for a discussion of the HD 189733 *Spitzer* transmission photometry and Beaulieu et al. (2010) for a detailed systematic removal comparison. We believe that in general the ramp removal is a solvable problem, meaning that the phenomenology of the ramp could be sufficiently well understood to resolve controversies that depend on small differences in the ramp function used in baseline fitting. As of this writing, we do not consider the ramp issue to be resolved.

### 4.2.2 HST/NICMOS

*HST*/NICMOS transmission spectrophotometry. A possible contradiction exists between water absorption seen using transit grism spectroscopy of HD189733b near 2 $\mu$m (Swain et al. 2008), and the same absorption in the same planet measured using multi-band filter photometry (Sing et al. 2009). The latter technique detects no water absorption, and is consistent with other *HST* measurements of atmospheric haze (Pont et al. 2008). However, the filter measurement and grism spectroscopy were well separated in time. Although we cannot rule out atmospheric variability (i.e., "weather") as a possible explanation for the discrepancy, the variability would be caused by a large change of about 5 scale heights over most of the limb of the planet (see also Sing et al. 2009). More work is needed to demonstrate that different *HST* instruments can produce mutually consistent contemporaneous measurements.

### 4.2.3 *HST*/STIS

Like other spectrographs, *HST*/STIS has increased correlated noise and reduced flux at the edges of the spectral orders. The uncertainties in the *HST*/STIS HD 209458 data set (Knutson et al. 2007b) may have been underestimated by as much as a factor of two (H. Knutson, private communication 2009). Any detections of spectral features at the edges of *HST*/STIS spectral orders are suspect and should be reinvestigated. Reported detections in the Knutson et al. (2007b) dataset include detection of water vapor (Barman 2007), of atomic H (Ballester, Sing, & Herbert 2007) and TiO and VO (Desert et al. 2008).

### 4.2.4 Ground-Based Measurements

Polarization of exoplanet atmospheres is an important phenomenon, because it can reveal the nature of atmospheric gases and aerosols. A tantalizing claimed detection of polarization from the hot Jupiter HD189733b (Berdygunia et al. 2008) reported polarization levels that exceeded theoretical expectations (Seager et al. 2000) by an uncomfortable margin. However, subsequent, independent work (Wiktorowicz, 2009) reports upper limits an order of magnitude smaller than the claimed detection, which is now believed to be erroneous.

### 4.2.5 A Comment on General Agreement

In contrast to the discrepancies we have discussed above, it is appropriate to point out the many cases where independent investigations have arrived at very similar conclusions. Usually very close agreement is found when different investigators analyze the same data. A case in point is GJ436b, where Deming et al. (2007), Gillon et al. (2007), and Demory et al. (2007) were in close accord as concerns the radius and 8 micron brightness temperature of that planet. In other instances, different investigators come to similar conclusions when analyzing different sets of independent data for different planets, an outstanding example being the prevalence of the inversion phenomenon. Whether or not it turns out to represent a genuine inversion of kinetic temperature, the prevalence of this phenomenon illustrates the conceptual consistency of the various Spitzer analyses of eclipsed hot Jupiters.

After our review on atmosphere discovery highlights (Section 3) and on observational challenges (section 4) we now turn to near future prospects.

## 5. Near Future Prospects

The observation and theory of exoplanets is now firmly established. Even though the current focus is on transiting hot Jupiter atmospheres, the future potential for exoplanet atmosphere studies is enormous. Here we outline the prospects for the next decade.

### 5.1 Direct Imaging of Giant Planets

The atmospheric study of young, hot Jupiters orbiting far from their stars is in flight as a new scientific frontier. Direct imaging has uncovered almost a dozen companion objects[1]. The most compelling objects are the three planets orbiting HR 8799 (Marois et al. 2008) and the planet orbiting Fomalhaut (Kalas et al. 2008). The HR 8799 planets orbit their host A star at 24, 38, and 68 AU respectively, and orbital motion over five years has been observed. Fomalhaut's planet has a mass constrained by the dust distortion dynamics. These objects are the most compelling planet *candidates* of the directly imaged stellar companions because most people agree they are planets; no case of three stars orbiting a central star has ever been seen. A concern with the planetary status of some of the other directly imaged substellar companions is that their mass is inferred from the object's brightness and age, via model planet evolutionary cooling tracks. The models used to infer mass have uncertainties especially with initial conditions that correspond to early times (Marley et al. 2007). The age of the host star is also a limiting factor for planet mass determination, and dominates for systems at modest ages (~100 Myr).

Spectra of younger and hotter objects—the objects that not everyone agrees are planets—are easier to obtain than for cooler objects. These hot objects (e.g., 1RXS J160929 in Lafreneier et al. 2008) resemble brown dwarfs much more than their cooler counterparts (e.g., 2M1207 and HR 8799) based on colors (Figure 13). Indeed, one of the most interesting near-term questions is to what extent do hot young giant planets resemble isolated brown dwarfs?

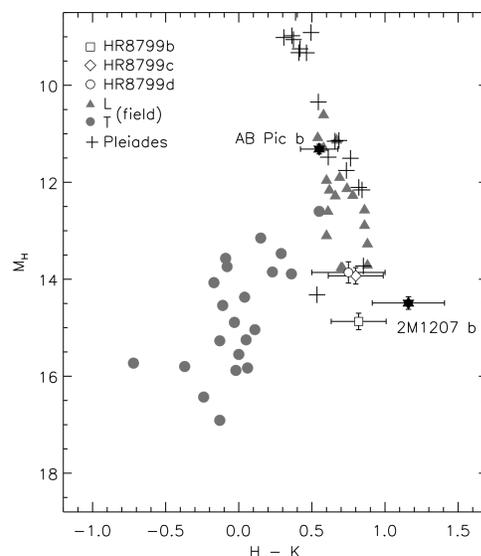

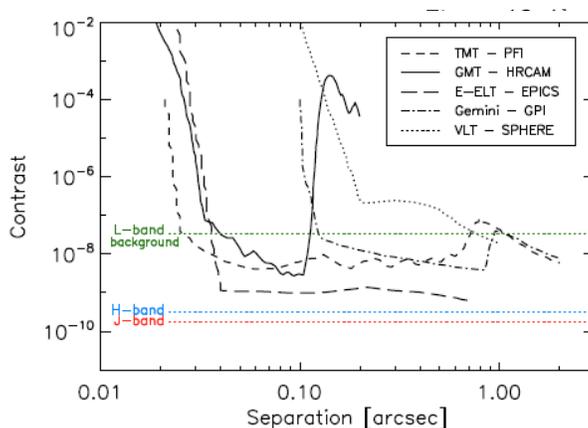

Figure 14. Direct imaging contrast vs. angular separation for next generation exoplanet imaging instruments. Contrasts corresponding to the sky background for one-hour integrations for the GMT on a Sun-like star at 10 pc is overplotted in color (dotted lines) for J, H, and L band, showing that observations will be background limited at longer wavelengths. The sensitivity curves for the TMT's planet imager PFI and Gemini's GPI are from (Macintosh et al. 2006) for a 4th magnitude target in H-band. The curve for 1-hour exposures of HRCAM on the GMT is also for H-band (GMT Conceptual Design Report: http://www.gmto.org/CoDRpublic). Corresponding sensitivity curves for SPHERE on VLT and EPICS on E-ELT (Kasper et al. 2008) are instead shown at lute magnitude in H-band versus H-K color. Old [...] and young Pleiades brown dwarfs (pluses) are [...] th two very low-mass brown dwarfs/planetary [...]ns (filled black symbols). Open symbols are HR [...], c (diamond), d (circle). The planet candidates [...]7b and HR8799 have very different colors than [...]fs. From Marois et al. (2008).

Another issue that is key to hot, young giant planet atmospheres is clouds. The physics of the transition between cloudy and clear spectra (i.e., the L to T transition for brown dwarfs) is not understood even for brown dwarfs (Helling et al. 2008)—it is instead modeled by fiat. Non-equilibrium chemistry is important to a lesser extent. These degeneracies appear strongly in colors (Figure 13). For example, HR 8799b colors can be fit to yield a huge range in effective temperature, $T$=900K to $T$=1500K depending on cloud parameters. Aside from the cloud complications, with future refined models and broad spectrophotometric coverage there is hope for determining an object's surface gravity by breaking degeneracies between composition and metallicity (although surface gravity for cooler objects is harder to separate from model uncertainties).

More atmosphere observations and interpretation should be forthcoming. With current telescopes and instrumentation, it is feasible to look at exoplanet-like objects down to temperatures in the 900 K range (e.g., 2M1207B, HR8799b). For the cooler objects, spectra at R ~50 SNR=10 from 1-2.4 microns are feasible using 8-10m telescopes and current-generation integral field spectrographs, particularly for wide-orbit planets. The next generation instrumentation, the Gemini Planet Imager and VLT SPHERE will allow spectral resolving power of R ~50 spectra of fainter planets, down to perhaps 300K. Higher-resolution spectra will require 20-30m ELTs. At those resolutions and higher SNR, compositional measurements become more feasible. See Figure 14 for an example of direct imaging capabilities. For a detailed review on high-contrast imaging with a focus on direct imaging of exoplanets, see Oppenheimer and Hinkley (2009) and Traub and Oppenheimer (2010).

## 5.2 Transiting Hot Jupiters and Hot Neptunes

Hot Jupiter atmosphere observations are expected to remain in the forefront in the near future not only because of their inherent favorable detection properties, but also because of new developments in ground-based observations. First, a population of very hot Jupiters have been discovered, with effective equilibrium temperatures of over 2000 K. These temperatures mean the planets have thermal emission at visible wavelengths and significant flux at near-IR wavelengths such that ground-based detections of secondary eclipses are possible (Lopez-Morales and Seager 2007). Indeed, a flurry of recent successful detections indicate the near-future fertility of ground-based secondary eclipse measurements. The near-infrared detections have been achieved from the ground; the first of these detections include a 6$\sigma$ detection in K-band of TrES-3b (with the WHT; de Mooij & Snellen 2009), a 4$\sigma$ detection in z'-band of OGLE-TR-56b (with Magellan and the VLT; Sing & Lopez-Morales 2009), and a 5$\sigma$ detection at 2.1 μm of CoRoT-1b (with the VLT Gillon et al. 2009).

The population of hot Jupiters tend to be around fainter stars than the bright "Rosetta Stones" HD 209458b and HD 189733b, because many hot Jupiters are discovered via transit searches that simultaneously monitor tens of thousands of faint stars. The fainter host stars actually help with ground-based observations because multiple comparison stars of similar brightness are generally more available than for bright stars, especially considering the limited field of view of large telescopes. A second enabling technical factor is telescope instrumentation that can be defocused and that have very fast readout rates. The photometry measurements can then achieve high duty-cycle and avoid saturation and the non-linear regime of the detectors (as well as avoid interpixel variations and flat-field errors).

Two things will happen with hot Jupiters. The opening of ground-based observations will enable observations of both many more target planets and a new wavelength range (at 2-5 μm). Visible wavelength observations combined with infrared data will help to understand the relative contribution of thermal emission and reflected light, and will enable constraints on the planetary albedo. Combined with Warm *Spitzer* observations and models, the new ground-based observations could enable constraints on the abundances of spectroscopically active gases and the presence and reflectivity of clouds or hazes that may exist in the atmospheres of these hot giant planets.

Space-based observations will complement ground-based hot Jupiter secondary eclipse observations.  The new space-borne instrumentation *HST*/COS (Froning & Greene 2009) will open the possibility of spectroscopic detection of molecular features in the UV to visible spectral range. As well, *HST*/WFC3 and the refurbished *HST*/STIS give renewed opportunities for detecting sodium as well as evidence for atmospheric escape. Secondary eclipses of *CoRoT* planets (Snellen, de Mooij, and Albrecht 2009) have been detected using visible light of the CoRoT data, and the *Kepler* team has reported a spectacular measurement of the eclipse of HAT-P-7b (Borucki et al. 2009).   Additional detections are expected from Kepler, and possibly from EPOXI (described in Ballard et al. 2009). On a further horizon, the eagerly awaited *JWST* (launch date 2014) will provide spectral resolution of about 2000, enabling even more detailed studies of transiting hot Jupiters.

## 5.3 Transiting Super Earths

### 5.3.1 Super Earth Atmospheres

In exoplanet research the frontier is always the most compelling. We are witnessing the gestation and birth of a new subfield of exoplanets: that of super Earths. Super Earths are unofficially defined as planets with masses between 1 and 10 Earth masses. The term super Earths is largely reserved for planets that are rocky in nature, rather than for planets with icy interiors or significant gas envelopes. The latter are often called mini-Neptunes. Because there may be a continuous and overlapping mass range between them, super Earths and mini-Neptunes are often discussed together.

Super Earths are a fascinating topic because they have no solar system counterparts, because they are our nearest term hope for finding habitable planets (Section 5.3.2), and because of their anticipated huge diversity (See Section 2.3).  The wide and almost continuous spread of exoplanet masses, semi-major axes and eccentricities of giant exoplanets illustrates the stochastic nature of planet formation and subsequent migration (Figure 1). This almost continuous distribution in parameter space surely extends to super Earths. Indeed, even though about 20 are known so far, all close to their stars (within 0.07 AU and one at 0.134 AU), their range of masses and orbits support this notion[1].

There is an exciting sense of anticipation in observing and studying super Earth atmospheres, because atmospheric formation and evolution is likely to yield a wide diversity. This is different from Jupiter and the other solar system gas giants, which have "primitive" atmospheres. That is, Jupiter has retained the gases it formed with, and these gases approximately represent the composition of the sun. The super Earth atmospheres, in contrast, could have a wide range of possibilities for the atmospheric mass and composition (see Figure 7). Attempts to evaluate these possibilities used calculations of atmospheres that formed by outgassing during planetary accretion, considering bulk compositions drawn from differentiated and/or primitive solar system meteoritic compositions (Elkins-Tanton & Seager 2008; Schaefer & Fegley 2010). Instead of narrowing down possibilities, this work emphasized the large range of possible atmospheric mass and composition of outgassed super Earths even before consideration of atmospheric escape.

Because of the huge range of possible parameter space for super Earth atmospheres, researchers take different paths, focusing on different regions of parameter space. One approach is to consider atmospheres similar to Earth, Venus or Mars (or their atmospheres in earlier epochs). Considering the amount of greenhouse gases including $CO_2$, Selsis et al. (2007) and von Bloh et al. (2007) both found that Gl 581d is more likely to be habitable (that is with surface temperatures consistent with liquid water) than Gl 581c. Other investigators consider atmospheres that radically depart from the terrestrial planets in our solar system. Water planets, akin to scaled up versions of Jupiter's icy moons, could have up to 50 percent water by mass, with concomitant massive steam atmospheres (Kuchner 2003; Leger et al. 2004; Rogers and Seager 2009). In a different approach, Miller-Ricci, Seager, and Sasselov (2009) considered GJ 581c and three possibilities relating to atmospheric hydrogen content. A suggestion of terrestrial planets with sulfur cycles dominating over carbon cycles is described in Kaltenegger and Sasselov (2010). Others have attempted to quantify the atmospheric escape of Earths and super Earths, with little success due to the unknown initial mass and star's activity history (e.g., Lammer et al. 2007).

The challenge in super Earth atmospheric research, compared to the initial hot Jupiter work, is that super Earths and their atmospheres occupy an almost unconstrained parameter space of mass and composition. Theory is currently leading observations, but observational success will be necessary to focus future efforts.

### 5.3.2 The M Star Opportunity for Habitable-Zone Super Earths

Habitable super Earths are the ones of most interest. All life on Earth requires liquid water, so a natural requirement in the search for habitable exoplanets is a planet with the right surface temperature to allow liquid water. Terrestrial-like planets are heated externally by the host star, so that a star's "habitable zone" is based on distance from the host star. Small stars have a habitable zone much closer to the star as compared to sun-like stars, owing to the lower luminosity of small stars compared to the sun.

The so-called M star opportunity is a boon for exoplanet research because the discovery and spectral characterization of a true Earth analog is immensely challenging, so is many years in the future. The search for super Earths orbiting a small star can accelerate our quest for a habitable world. The renewed interest in M-dwarf stars is described in Scalo et al. (2007) and Tarter et al. (2007), and observational rationale is articulated in Charbonneau and Deming (2007). Observational searches include radial velocity with transit followup (Udry & Santos 2007) and targeted star searches for transits (Nutzmann & Charbonneau 2007; Charbonneau et al. 2009).

The chance of discovering a habitable-zone super Earth transiting a low-mass star in the immediate future is tremendously high. Indeed, the recently detected super Earth orbiting GJ1214 (Charbonneau et al. 2009) is already close to a habitable temperature, at $\sim$ 450K. Observational selection effects favor the discovery of super Earths orbiting in the habitable zones of M stars compared to Earth analogs in almost every way. The magnitude of the planet transit signature is related to the planet-to-star area ratio. Low-mass stars can be 2 to 10 times smaller than a sun-like star, improving the total transit depth signal from about 1/10000 for an Earth transiting a sun, to 1/2500 or 1/100 for the same-sized planet. A planet's equilibrium temperature scales as $T_{eq} \sim T_* \, (R_*/a)^{1/2}$, where $a$ is the planet's semi-major axis. The temperature of low mass stars is about 3500 to 2800 K (for stars smaller than the sun by 2 and 10 times, respectively). The habitable zone of a low-mass star would therefore be 4.5 to 42 times closer to the star compared to the sun's habitable zone. To measure a planet mass, the radial velocity semi-amplitude, $K$, scales as $K \sim (a \, M_*)^{-1/2}$, and the low-mass star masses are 0.4 to 0.06 times that of the sun. Obtaining a planet mass is therefore about 3 to 30 times more favorable for an Earth-mass planet orbiting a low-mass star compared to an Earth-sun analog. The transit probability scales as $R_*/a$. The probability for a planet to transit in a low-mass star's habitable zone is about 2.3% to 5%, much higher as compared to the low 0.5% probability of an Earth-sun analog transit. Finally, from Kepler's Third Law, a planet's orbital period, $P$, scales as $P \sim a^{3/2}/M_*^{1/2}$, meaning that the period of a planet in the habitable zone of a low-mass star is 7 to 90 times shorter than the Earth's 1 year period, and the planet transit can be observed often enough to build up a signal (as compared to an Earth analog once a year transit). A super Earth larger than Earth (and up to about 10 $M_\oplus$ and 2 $R_\oplus$) is even easier to detect by its larger transit signal and mass signature than Earth.

Debate on how habitable a planet orbiting close to an M star can actually be is an active topic of research (see the reviews by Scalo et al. 2007; Tarter et al. 2007). Some previously accepted "show stoppers" are no longer considered serious impediments. Atmospheric collapse due to cold night-side temperatures on a tidally-locked planet will not happen as long as the atmosphere is thick enough (0.1 bar) for atmospheric circulation to redistribute absorbed stellar energy to heat the night side (Joshi, Haberle,

& Reynolds 1997). Short-term stellar variability due to large amplitude star spots could change the planet's surface temperature by up to 30 K in the most severe cases (Joshi et al. 1997; Scalo et al. 2007); but even some terrestrial life can adapt to freeze-thaw scenarios. Bursts of UV radiation due to stellar flares could be detrimental for life, but the planet's surface could be protected by a thick abiotic ozone layer (Segura et al. 2005), or alternatively life could survive by inhabiting only the subsurface.

Other concerns about habitability of planets orbiting close to M stars have not yet been resolved. Flares and UV radiation could erode planet atmospheres (Lammer et al. 2007), especially because the active phase of M stars can last for billions of years (West et al. 2008). Tidally-locked planets in M star habitable zones will be slow rotators; a weak magnetic field due to an expected small dynamo effect will not protect the atmosphere from erosion. Planets accreting in regions that become habitable zones of M dwarf stars form rapidly (within several million years); the planet may not have time to accrete volatiles (e.g., water) that are present in the protoplanetary disk much farther away from the star (Lissauer 2007).

Observers will never limit their plans based of theoretical reasoning, and so the search for life on planets in the habitable zones of M stars will proceed.

### 5.3.3 JWST and Transiting Super Earths Observations

We anticipate the discovery of a handful of rare but highly valuable transiting super Earths in the habitable zones of the brightest low-mass stars. With such prize targets, astronomers will strive to observe the transiting super Earth atmospheres in the same way we are currently observing transiting hot Jupiters orbiting sun-like stars.

Fortunately, the nearest example of a habitable transiting super Earth is expected to lie closer than 35 parsecs (Deming et al. 2009). If such a system can be found using an all-sky survey like TESS (Ricker et al. 2010), NASA's *JWST* scheduled for launch in 2014, will be capable of observing the absorption signatures of major molecules like water and carbon dioxide. Such observations will require monitoring of multiple transits, often amounting to ~ 100 hours of *JWST* observation. For example, in our simulations (Deming et al. 2009), one habitable Super Earth lying at 22 parsecs distance required 85 hours of *JWST*/NIRSpec observations to measure carbon dioxide absorption during transit to S/N = 28. The atmospheres of super Earths with temperatures above the habitable range will be observable by *JWST* using only a few transits and/or eclipses. For a comparison of Earth-like planet atmospheres around both M stars and Sun-like stars see Kaltenegger & Traub (2009).

Observations for planets with habitable temperatures will be challenging, due to the thinner atmospheres and lower scale heights on cooler rocky planets compared to the puffy atmospheres of hot Jupiter (gas giant) planets. In other words, using transmission spectroscopy during transit, atmospheres with large scale heights are easier to detect than are atmospheres with small scale heights (Miller-Ricci et al. 2008). (The scale height is defined by $H = kT/\mu_m g$, where $k$ is Boltzmann's constant, $T$ is temperature, $\mu_m$ is mean molecular weight, and $g$ is surface gravity.) Strong molecular bands can be optically thick in transmission over many scale heights, so atmospheres that stretch this absorption over the maximum height range (i.e., have large scale heights) will block the most star light, and will produce maximum absorption in transmission spectra. The scale height formula explains why transmission spectra observations favor hot, or hydrogen-rich atmospheres on planets with low surface gravities and not, in contrast, planets with high bulk density, having thin atmospheres of high molecular weight. Despite the difficulty of detecting $CO_2$ atmospheres in transmission, and emission we have shown that it is feasible with favorable target stars (Deming et al. 2009).

## 6. Future Detection and Characterization of Earth Analogs

Without question the holy grail of exoplanet research is the discovery of a true Earth analog: an Earth-size, Earth-mass planet in an Earth-like orbit about a sun-like star. This is in spite of: the near-term obsession on super Earths that are tidally-locked in the habitable zones of M stars; the awesome diversity of exoplanets; and the fact that if an Earth-like planet can be detected by a given telescope, so too can a super Earth. Humanity will always have a strong desire to search for a place like home.

Discovery of an Earth analog is a massive challenge for each of the different exoplanet discovery techniques (Figure 1) because Earth is so much smaller (~1/100 in radius), so much less massive (~$1/10^6$), and so much fainter (~$10^7$ for mid-IR wavelengths to ~$10^{10}$ for visible wavelengths) than the sun. Exoplanet discovery techniques other than direct imaging may have an easier time in *finding* an Earth-mass planet or Earth-size planet. These other methods include space-based astrometry (see Shao & Nemati 2009 and references therein) or potentially ground-based radial velocity with the new development of the astro frequency comb spectrograph (Steinmetz et al. 2008; Li et al. 2009). Additionally, NASA's *Kepler Space Telescope* is using the transit technique to take a census of Earth-size, Earth-like transiting planets orbiting sun-like stars to tell us how common Earth-size planets in Earth-like orbits actually are (Borucki et al. 2010).

We emphasize that discovery of Earth-size or Earth-mass planets—even those in their star's habitable zone—is not the same as identifying a habitable planet. Venus and Earth are both about the same size and mass—and would appear the same to an astrometry, radial-velocity, or transit observation. Yet Venus is completely hostile to life due to the strong greenhouse effect and resulting high surface temperatures (over 700 K), while Earth has the right surface temperature for liquid water oceans and is teeming with life. This is why, in the search for habitable planets, we must hold on to the dream of a direct-imaging space-based telescope capable of blocking out the starlight.

## 6.1 Earth as an Exoplanet

Earth's atmosphere is a natural starting point when considering planets that may have the right conditions for life or planets that have signs of life in their atmospheres. Earth from afar, with any reasonably sized space telescope, would appear only as a point of light, without spatial resolution of surface features. We have real atmospheric spectra of the hemispherically-integrated Earth (Figure 15) by way of Earthshine measurements at visible and near-IR wavelengths (e.g., Turnbull et al. 2007) and from spacecraft that turn to look at Earth (e.g., Pearl and Christensen 1997). Earth has several strong spectral features that are uniquely related to the existence of life or habitability. The gas oxygen ($O_2$) makes up 21 percent of Earth's atmosphere by volume, yet $O_2$ is highly reactive and therefore will remain in significant quantities in the atmosphere only if it is continually produced. On Earth plants and photosynthetic bacteria generate oxygen as a metabolic byproduct; there are no abiotic continuous sources of large quantities of $O_2$. Ozone ($O_3$) is a photolytic product of $O_2$, generated after $O_2$ is split up by the sun's UV radiation. Oxygen and ozone are Earth's two most robust biosignature gases (Leger et al. 1993). Nitrous oxide ($N_2O$) is a second gas produced by life—albeit in small quantities—during microbial oxidation-reduction reactions.

Other spectral features, although not biosignatures because they do not reveal direct information about life or habitability, can nonetheless provide significant information about the planet. These include $CO_2$ (which is indicative of a terrestrial atmosphere and has an extremely strong mid-infrared spectral feature) and $CH_4$ (which has both biotic and abiotic origins).

In addition to atmospheric biosignatures the Earth has one very strong and very intriguing biosignature on its surface: vegetation. The reflection spectrum of photosynthetic vegetation has a dramatic sudden rise in albedo around 750 nm by almost an order of magnitude. Vegetation has evolved this strong reflection feature, known as the "red edge", as a cooling mechanism to prevent overheating which would cause chlorophyll to degrade (e.g., Seager et al. 2005, Kiang et al. 2007.) On Earth this signature is reduced to a few percent (see, e.g. Montanes-Rodriguez et al. 2006; Arnold 2008; and references therein) because of leaf canopy structure, continental fraction, and cloud coverage over forested areas. Such a spectral surface feature could, however, be much stronger on a planet with a lower cloud cover fraction. Recall that any observations of Earth-like exoplanets will not be able to spatially resolve the surface. A surface biosignature could be distinguished from an atmospheric signature by time variation; as the continents, or different concentrations of the surface biosignature, rotate in and out of view the spectral signal will change correspondingly.

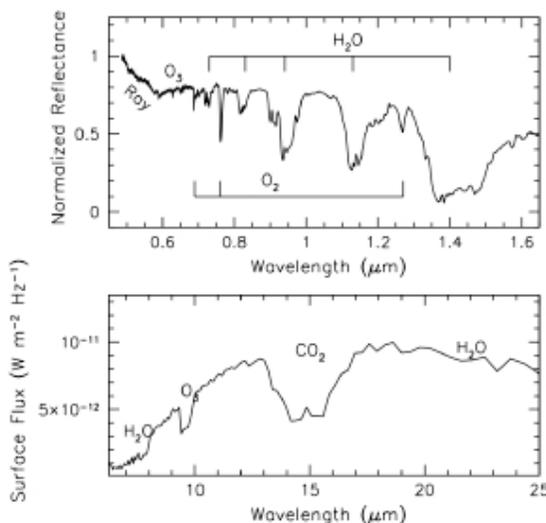

Figure 15. Earth's hemispherically averaged spectrum. Top: Earth's visible and near-IR wavelength spectrum from Earthshine measurements (Turnbull et al. 2006). Bottom: Earth's mid-infrared spectrum as observed by Mars Global Surveyor enroute to Mars (Pearl and Christenson 1997). Major molecular absorption features are noted; Ray means Rayleigh scattering.

Earth viewed from afar would also vary in brightness with time, due to the brightness contrast of cloud, land, and oceans. As Earth rotates and continents come in and out of view, the total amount of reflected sunlight will change due to the high albedo contrast of different components of Earth's surface (< 10% for ocean, > 30-40% for land, > 60% for snow and some types of ice). In the absence of clouds this variation could be an easily detectable factor of a few. With clouds the variation is muted to 10 to 20% (Ford, Seager, and Turner 2001). From continuous observations of Earth over a few month period, Earth's rotation rate could be extracted, weather identified, and the presence of continents inferred. Palle et al. (2008) modeled Earth as an exoplanet using three months of cloud data taken from satellite observations. They showed that a hypothetical distant observer of Earth could measure Earth's rotation rate. This is surprising and means that, despite Earth's dynamic weather patterns, Earth has a relatively stable signature of cloud patterns. These cloud patterns arise in part because of Earth's continental arrangement and ocean currents. Beyond detecting Earth's rotation rate, Palle et al. (2008) found deviations from the periodic photometric signal, indicative to hypothetical distant observers that active weather is present on Earth.

Real data of the spatially unresolved Earth is available. Global, instantaneous spectra and photometry can be obtained from observations from Earth itself—by Earthshine measurements (Arnold et al. 2002; Woolf et al. 2002). Earthshine, easily seen with the naked eye during crescent moon phase (Figure 15, top panel), is sunlight scattered from Earth that scatters off of the moon and travels back to Earth. Earthshine data is more relevant to studying Earth as an exoplanet than remote sensing satellite data. The latter is highly spatially resolved and limited to narrow spectral regions. Furthermore by looking straight down at specific regions of Earth, hemispherical flux integration with lines-of-sight through different atmospheric path lengths is not available.

Recently, the NASA *EPOXI* spacecraft viewed Earth from afar—31 million miles distant. *EPOXI* is a NASA Discovery Mission of Opportunity, formerly the Deep Impact mission that impacted and observed Comet Temple. *EPOXI* spent several months in 2008-2009 observing stars with known exoplanets as well as observing Earth as an exoplanet. The *EPOXI* spacecraft obtained light curves of Earth at seven wavebands spanning 300-1000 nm. Using multi-wavelength observations each spanning one day, Cowan et al. (2009) found that the rotation of Earth leads to diurnal albedo variations of 15-30%, with the largest relative changes occurring at the reddest wavelengths. Using a principal component analysis of the multi-band light curves, Cowan et al. (2009) found that 98% of the diurnal color changes of Earth are due to only two dominant eigencolors. The spectral and spatial distributions of the eigencolors correspond to cloud-free continents and oceans, enabling construction of a crude longitudinally averaged map of Earth.

Beyond characterization of light curves and spectra, the search for other Earths is also a search for biosignature gases.

## 6.2 Biosignature Gases

An atmospheric biosignature gas is one produced by life. Life metabolizes and generates metabolic byproducts. Some metabolic byproducts dissipate into the atmosphere and can accumulate as biosignature gases. In exoplanets, then, we focus on a "top down" approach of a biosignature framework. In the top down approach, we do not worry about what life is, just on what life does (i.e., life metabolizes). The "bottom up" approach—the details of the origins and evolution of life is left to the biologists.

The canonical concept for the search for atmospheric biosignatures is to find an atmosphere severely out of thermochemical redox equilibrium (Lederberg 1965; Lovelock 1965). Redox chemistry adds or removes electrons from an atom or molecule (reduction or oxidation, respectively). Redox chemistry is used by all life on Earth and is thought to enable more flexibility than non-redox chemistry. The idea is that gas byproducts from metabolic redox reactions can accumulate in the atmosphere and would be recognized as biosignatures because abiotic processes are unlikely to create a redox disequilibrium. Indeed Earth's atmosphere has oxygen (a highly oxidized species) and methane (a very reduced species) several orders of magnitude out of thermochemical redox equilibrium.

In practice it could be difficult to detect both molecular features of a redox disequilibrium pair. The Earth as an exoplanet, for example (Figure 15), has a relatively prominent oxygen absorption feature at 0.76 $\mu$m, whereas methane at present-day levels of 1.6 ppm has only extremely weak spectral features. During early Earth $CH_4$ may have been present at much higher levels (1000 ppm or even 1%), as possibly produced by wide-spread methanogen bacteria (Haqq-Misra et al. 2008 and references therein). Such high $CH_4$ concentrations would be easier to detect, but since the Earth was not oxygenated during early times the $O_2$-$CH_4$ redox pairs would not be detectable concurrently. (See Des Marais et al. 2001.)

The more realistic atmospheric biosignature gas is a single gas completely out of chemical equilibrium. Earth's example again is oxygen or ozone, about eight orders of magnitude higher than expected from equilibrium chemistry and with no known abiotic production at such high levels. The challenge with a single biosignature outside of the context of redox chemistry becomes one of false positives. To avoid false positives we must look at the whole atmospheric context. For example, a high atmospheric oxygen content might indicate a planet undergoing a runaway greenhouse with evaporating oceans. When water vapor in the atmosphere is being photodissociated with H escaping to space, $O_2$ will build up in the atmosphere for a short period of time. In this case, $O_2$ can be associated with a runaway greenhouse via very saturated water vapor features, since the atmosphere would be filled with water vapor at all altitudes. Other $O_2$ and $O_3$ false positive scenarios (Selsis et al. 2002) are discussed and countered in Segura et al. (2007).

Most work to date has focused on mild extensions of exoplanet biosignatures as on Earth ($O_2$, $O_3$, $N_2O$) or early Earth (possibly $CH_4$) biosignatures. Research forays into biosignature gases that are negligible on Earth but may play a more dominant role on other planets has started. Pilcher (2003) suggested that organosulfur compounds, particularly methanethiol ($CH_3SH$, the sulfur analog of methanol) could be produced in high enough abundance by bacteria, possibly creating a biosignature on other planets. Pilcher (2003) emphasized a potential ambiguity in interpreting the 9.6 $\mu$m $O_3$ spectral feature since a $CH_3SH$ feature overlaps with it. Segura et al. (2005) showed that the Earth-like biosignature gases $CH_4$, $N_2O$, and even $CH_3Cl$ have higher concentrations and therefore stronger spectral features on planets orbiting M stars compared to Earth. The reduced UV radiation on quiet M stars enables longer biosignature gas lifetimes and therefore higher concentrations to accumulate. Seager and Schrenk (2010) have reviewed Earth-based metabolism to summarize the range of gases and solids produced by life on Earth. A fruitful new area of research will be on which molecules are potential biosignatures and which can be identified as such on super Earth planets different from Earth.

Breaking free from terracentrism, the NRC Report on "The Limits of Organic Life in Planetary Systems" by Baross et al. (2007), proposed that the conservative requirements for life of liquid water and carbon could be replaced by the more general requirements of a liquid environment and an environment that can support covalent bonds (especially between hydrogen, carbon, and other atoms). This potentially opens up a new range for habitable planets, namely those beyond the ice line in a cryogenic habitable zone where water is frozen but other liquids such as methane and ethane are present. Although no one has yet studied the possibility of exoplanet biosignatures on cold exoplanets, research on what it takes for life in non-water liquids (Bains 2004) and the possibilities for methanogenic life in liquid methane on the surface of Titan (McKay and Smith 2005) are a useful start.

Biosignature research is only just beginning to unfold and seems to be fertile ground for new lines of investigation—especially with the considerable interest it draws. Nonetheless we will always be humbled by the warning that in many cases we will not be 100 percent sure of a biosignatures definite attribution to life. Before we can become too serious about any future detection of biosignatures, however, we must visit how we can detect Earth-like planets that would support life.

### 6.3 Future: Detection Prospects for Earth Analogs

Discovery of an Earth analog is a massive challenge for each of the different exoplanet discovery techniques (Figures 1 and 4) and direct imaging is no exception. In order to observe a planet under the very low Earth-sun flux contrast, at visible wavelengths the diffracted light from the star must be suppressed by 10 billion times. This level of suppression is not possible from ground-based telescopes, even the ELT of the future, because no source is bright enough for adaptive optics; a guide star of magnitude $R = -7$ would be needed (Traub and Oppenheimer 2010). Space-based direct imaging is the best option for detection and characterization of Earth analogs.

A collection of related ideas to supress the starlight are called "coronagraphs" a term that originated with artificial blocking out sunlight to observe the sun's corona (Lyot 1939). Novel-shaped apertures, pupils, and pupil masks to suppress some or all of the diffracted star light have been developed in the last few years (see, e.g.,Trauger and Traub 2007 and references therein). At mid-infrared wavelengths, an interferometer would be needed to avoid the exceedingly large aperture to spatially separate an Earth-like planet from its star for moderate distances from Earth (i.e., because of the $\lambda/D$ scaling). See Cockell et al. (2009) and references there in for a description of the Darwin mission. At the present time, there are no plans to build and launch a Terrestrial Planet Finder or Darwin mission. The price tag and complexity are almost prohibitive.

A recent development has given renewed promise for a Terrestrial Planet Finder type of mission. The idea is to use the already planned *JWST* (launch date 2014) together with a novel-shaped external occulter placed tens of thousands of kilometers from the telescope in order to suppress the diffracted starlight (Cash 2006; Soummer et al. 2009). Most of the time the *JWST* would be functioning as planned, and during this time the starshade would fly across the sky to the next target star. While there are many technical and programmatic concerns for the *JWST* + external occulter idea, none are without solution.

## 7. Concluding Remarks

The field of exoplanet atmospheres is firmly established with a set of hot Jupiter inaugural observations and interpretation as the foundation. We see a cycle for atmospheric studies (Figure 16) that begins with predictions at a time when observations are leading theory. Next in the cycle comes the first truly breakthrough observations that enable a flurry of further observations. Third comes a period of interpretation or perhaps more aptly termed *retrodiction*: modeling and theory work that may raise more questions than answers and beg for better data; this part of the cycle is where observations are leading theory. We are at this point with hot Jupiter atmospheres. Closing the cycle comes when higher S/N and higher spectral resolution data become available to answer the outstanding questions and provide closure. We foresee this closure for transiting hot Jupiters with future *JWST* observations.

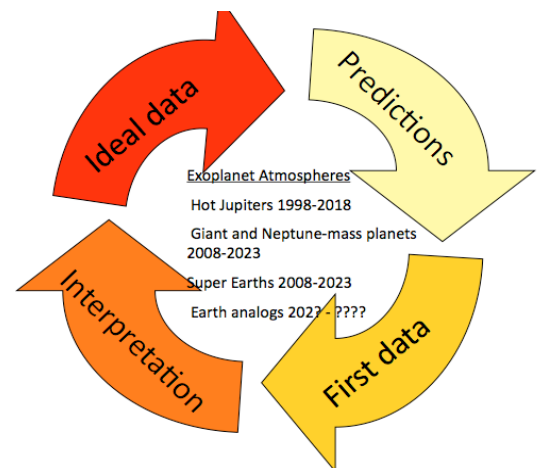

Figure 16. Cycles for exoplanet atmospheres.

With the cycle picture in mind we envision four eras for exoplanet atmospheric studies. The first is the hot Jupiter studies, with the start of the cycle in 1998 and with at least some closure by five years after *JWST* launch. The second era is that of more orbitally-distant giant planets and Neptunes, beginning now with direct imaging of young, hot Jupiters far from their stars, maturing with the next generation Gemini and VLT instrumentation, and finding some closure with the very large ground-based telescopes TMT/GMT/ELT of the future. The third era is also just beginning, that of transiting super Earths and mini Neptunes. With large amounts of *JWST* time the era of super Earths will advance to the third step of the exoplanet atmospheric cycle. The fourth era is that of true Earth analogs. (Once technology can reach Earth analog planet atmospheres, almost any kind of larger planet's or moon's atmosphere can also be studied.) Predictions using Earth as an exoplanet and some extensions are underway, but the first observations will have to await a specialized space telescope that can block out the orders of magnitude brighter starlight.

At a few special times in history, astronomy changed the way we see the Universe. Hundreds of years ago, humanity believed that Earth was the center of everything—that the known planets and stars all revolved around Earth. In the late 16th century, the Polish astronomer Nicolaus Copernicus presented his revolutionary new view of the Universe, where the sun was the center, and Earth and the other planets all revolved around it. Gradually, science adopted this "Copernican" theory (solidly after Comet Halley's successfully predicted return), but this was only the beginning. In the early 20th century, astronomers concluded that there are galaxies other than our own Milky Way. Astronomers eventually recognized that our Sun is but one of hundreds of billions of stars in our galaxy, and that our galaxy is but one of upwards of hundreds of billions of galaxies. When and if we find that other Earths are common and see that some of them have signs of life, we will at last complete the Copernican Revolution—a final conceptual move of the Earth, and humanity, away from the center of the Universe. This is the promise and hope for exoplanet atmospheres—the detection and characterization of habitable worlds.

## Acknowledgements
That the field of exoplanet atmospheres has matured enough for an ARAA chapter dedicated to exoplanet atmospheres is a tribute to all who have worked hard to get where we are today. A special thank you to our colleagues who invested critical time and effort in the earliest years when rewards and data were limited. We apologize in advance to all researchers whose work could not be appropriately cited due to space limitations. We thank Bruce Macintosh, Nikku Madhusudhan, Heather Knutson, Jean-Michel Desert, Bryce Croll, Frederic Pont, and Adam Showman for useful discussions.

We thank the *Spitzer* SSC and all of the people who worked on *Spitzer* and *HST* from concept through launch and operations.